\documentclass[aps,prd,twocolumn,showpacs,preprintnumbers,superscriptaddress]{revtex4}

\usepackage{graphicx}
\usepackage{amssymb}
\usepackage{amsmath}
\usepackage{subfigure}
\usepackage{dcolumn}
\usepackage{bm}
\usepackage{natbib}
\usepackage{multirow}
\usepackage{aas_macros}
\usepackage{enumerate}

\begin{document}

\newcommand{\rem}[1]{{\bf #1}}

\preprint{ICRR-Report-581-2010-14,\ IPMU11-0032}

\title{Study of gravitational radiation from cosmic domain walls}

\author{Masahiro Kawasaki}
\email{kawasaki@icrr.u-tokyo.ac.jp}
\affiliation{Institute for Cosmic Ray Research, The University of Tokyo, 
5-1-5 Kashiwa-no-ha, Kashiwa City, Chiba 277-8582, Japan} 
\affiliation{Institute for Physics and Mathematics of the Universe, 
The University of Tokyo,
5-1-5 Kashiwa-no-ha, Kashiwa City, Chiba 277-8582, Japan}
\author{Ken'ichi Saikawa}
\email{saikawa@icrr.u-tokyo.ac.jp}
\affiliation{Institute for Cosmic Ray Research, The University of Tokyo, 
5-1-5 Kashiwa-no-ha, Kashiwa City, Chiba 277-8582, Japan} 

\date{\today}

\begin{abstract}
In this paper, following the previous study, we evaluate the spectrum of gravitational wave background generated by domain walls
which are produced if some discrete symmetry is spontaneously broken in the early universe.
We apply two methods to calculate the gravitational wave spectrum:
One is to calculate the gravitational wave spectrum directly from numerical simulations,
and another is to calculate it indirectly by estimating the unequal time anisotropic stress power spectrum of the scalar field.
Both analysises indicate that the slope of the spectrum changes at two characteristic frequencies corresponding to the
Hubble radius at the decay of domain walls and the width of domain walls, and that the spectrum between these two characteristic
frequencies becomes flat or slightly red tilted. The second method enables us to evaluate the GW spectrum
for the frequencies which cannot be resolved in the finite box lattice simulations, but relies on the assumptions for
the unequal time correlations of the source.
\end{abstract}


\maketitle

%%%%%%%%%%%%%%%%%%%%%%%%%%%%%%%%%%%%%%%%%%%%%%%%%%%%%%%%%%%%%%%%%%%%%%
\section{\label{sec1}Introduction}
%%%%%%%%%%%%%%%%%%%%%%%%%%%%%%%%%%%%%%%%%%%%%%%%%%%%%%%%%%%%%%%%%%%%%%
Gravitational wave (GW) is one of the robust predictions of general relativity, and expected to be detected in the next decades.
Since GWs have few interactions with matter and radiation, they propagate almost freely after their production.
Therefore, analogously to the cosmic microwave background,
the search for a stochastic background of GWs will give us rich informations about the early universe which has not been probed by electromagnetic waves. 
Various mechanisms to generate GW backgrounds have been proposed, such as inflation~\cite{2006PhRvD..73b3504S},
(p)reheating after inflation~\cite{1997PhRvD..56..653K,2006JCAP...04..010E,2007PhRvL..98f1302G,2007PhRvD..76l3517D,2009JCAP...03..001D}, 
cosmic strings~\cite{2005PhRvD..71f3510D,2007PhRvL..98k1101S},
and first order phase transitions~\cite{1992PhRvD..45.4514K,1992PhRvL..69.2026K,1993PhRvD..47.4372K,1994PhRvD..49.2837K}.
In addition to them, we propose that domain walls, which are surface like topological defects produced
when a discrete symmetry is spontaneously broken, can be another source of the stochastic background of GWs.
The existence of domain walls is cosmologically unacceptable, since they eventually overclose the universe~\cite{1974JETP...40....1Z}.
However, if domain walls are unstable and decay at a sufficiently early time~\cite{1981PhRvD..23..852V,1989PhRvD..39.1558G},
the energy stored in them would be radiated as GWs,
which become the stochastic background observed today~\cite{1998PhRvL..81.5497G,2010JCAP...05..032H}.
There are several particle physics models which predict such phenomena.
For instance, the spontaneous breaking of the discrete R symmetry in the theory with supersymmetry induces domain walls
which decay when the Hubble parameter becomes comparable to the scale of the gravitino mass~\cite{2008PhLB..664..194T,2010JHEP...07..003D}.
Also, the theory of axions, which was introduced in order to solve the strong CP problem of quantum chromodynamics,
naturally predicts the existence of domain walls~\cite{1999PhRvD..59b3505C,2010arXiv1012.4558H}.
Therefore, the observation of the stochastic background of GWs produced by domain walls
can become another probe of the theory beyond the standard model of particle physics.

The stochastic background of GWs of cosmological origin is expected to be
isotropic, stationary and unpolarized, and therefore characterized by its frequency spectrum \cite{2000PhR...331..283M}.
In the previous study~\cite{2010JCAP...05..032H}, we calculated the spectrum of GWs produced by domain walls
based on the numerical simulation of the scalar field in the expanding universe.
The results were straightforwardly obtained, and it was shown that GWs from domain walls have a broad and nearly flat spectrum.
However, in the previous work, we chose the energy scale of the symmetry breaking as an unrealistic value $\eta\simeq 10^{17}$GeV in order to
follow the domain wall evolution from the initial thermal state, 
and simply extrapolated the numerical result to predict the spectrum observed today.
This estimation gives the large uncertainty, which can be a factor of ${\cal O}(10^{\pm 1})$ in the magnitude of GWs.
Furthermore, we put the thermal initial condition for numerical simulations, which made an additional peak at high frequencies in the GW spectrum.
This might contaminate the spectrum of GWs purely generated from domain walls.
In order to remove these difficulties and give more accurate predictions for future observations,
it is necessary to investigate further about the spectrum of GWs.

In this work, we apply two approach to evaluate the spectrum of GWs produced by domain walls.
First, we perform three dimensional lattice simulation of domain walls, and calculate the GW spectrum
directly from the results of the numerical simulations. This procedure is similar to that in our previous work, but
we work with different setups: While we performed the calculations only for radiation dominated background in the previous work,
we calculate the GW spectrum both in radiation and matter dominated backgrounds.
Furthermore, we do not assume the coupling with the thermal bath. In this case, the GW spectrum
would represent the feature of that purely produced by domain walls.
This analysis will confirm and clarify the features of the GW spectrum which we found in the previous work.
Second, we reevaluate the GW spectrum indirectly by using the anisotropic stress power spectrum obtained from numerical simulations.
For the evaluation of the GW spectrum, we introduce the same approximations which has been used to calculate the spectrum of GWs
generated by first order phase transitions~\cite{2008PhRvD..77l4015C,2009PhRvD..79h3519C,2009JCAP...12..024C}.
This method depends on various assumptions about the source of GWs, but it might become an alternative way which
enables us to evaluate the GW spectrum without relying on numerical simulations with a long dynamical range.

This paper is organized as follows. In section~\ref{sec2}, we describe a notation to calculate GWs and derive a formula which gives
the spectrum of GWs produced in the arbitrarily expanding background.
Then, we present the results of numerical simulations and discuss the features of the spectrum of GWs produced by domain walls in section~\ref{sec3}.
In section~\ref{sec4}, we take another approach and reevaluate the GW spectrum. We also give forecasts for future observations
based on the result of the analysis performed there.
Finally we conclude in section~\ref{sec5}.

%%%%%%%%%%%%%%%%%%%%%%%%%%%%%%%%%%%%%%%%%%%%%%%%%%%%%%%%%%%%%%%%%%%%%%
\section{\label{sec2}A stochastic gravitational wave background in the expanding universe}
%%%%%%%%%%%%%%%%%%%%%%%%%%%%%%%%%%%%%%%%%%%%%%%%%%%%%%%%%%%%%%%%%%%%%%
In this section, we derive basic equations for GWs produced in arbitrary expanding background.
We consider a spatially flat Friedmann-Robertson-Walker background in which GWs are represented
by the spatial metric perturbation
\begin{equation}
ds^2 = -dt^2+a^2(t)(\delta_{ij}+h_{ij})dx^idx^j, \label{eq2-1}
\end{equation}
where $h_{ij}$ satisfies the transverse-traceless (TT) condition $\partial_ih_{ij}=h^i_i=0$.
Since we will investigate the generation of GWs in both radiation and matter dominated background,
it is convenient to consider the arbitrarily expanding background where the scale factor evolves as
\begin{equation}
a(t) \propto \tau^{\alpha} \propto t^{\beta}, \label{eq2-2}
\end{equation}
where $\tau$ is conformal time defined by $d\tau=dt/a$. 

The metric perturbations $h_{ij}$ obey the linearized Einstein equation
\begin{eqnarray}
\ddot{h}_{ij}(t,{\bf x}) + 3H\dot{h}_{ij}(t,{\bf x}) - \frac{\nabla^2}{a^2}h_{ij}(t,{\bf x})\nonumber \\ 
= \frac{16\pi G}{a^2}T^{\mathrm{TT}}_{ij}(t,{\bf x}), \label{eq2-3}
\end{eqnarray}
where a dot denotes a derivative with respect to cosmic time $t$ and $T^{\mathrm{TT}}_{ij}$ is the TT part of the stress-energy tensor.
If we work in spatial Fourier space and change time variable from cosmic time $t$ into conformal time $\tau$, this equation gives
\begin{eqnarray}
h''_{ij}(\tau,{\bf x}) + \frac{2\alpha}{\tau}h'_{ij}(\tau,{\bf k}) + k^2h_{ij}(\tau,{\bf k})\nonumber \\ 
 = 16\pi GT^{\mathrm{TT}}_{ij}(\tau,{\bf k}), \label{eq2-4}
\end{eqnarray}
where a prime denotes a derivative with respect to conformal time $\tau$. Defining the rescaled metric
\begin{equation}
\bar{h}_{ij} = ah_{ij}, \label{eq2-5}
\end{equation}
we obtain
\begin{eqnarray}
\left[\frac{\partial^2}{\partial x^2} + \left(1-\frac{4\nu^2-1}{4x^2}\right)\right]\bar{h}_{ij}(\tau,{\bf k})\nonumber \\ 
= \frac{16\pi Ga(\tau)}{k^2}T^{\mathrm{TT}}_{ij}(\tau,{\bf k}), \label{eq2-6}
\end{eqnarray}
where $x=k\tau$, and $\nu$ is defined by
\begin{equation}
\nu = \alpha - \frac{1}{2} = \frac{3\beta-1}{2(1-\beta)}. \label{eq2-7}
\end{equation}

Let us assume that the source term $T^{\mathrm{TT}}_{ij}$ is nonzero during the interval $\tau_i\le\tau\le\tau_f$.
The solution of eq.~(\ref{eq2-6}) with initial conditions $\bar{h}_{ij}(\tau_i)=\bar{h}'_{ij}(\tau_i)=0$ is given by the
time integral of the source term convoluted with a Green function
\begin{widetext}
\begin{align}
\bar{h}_{ij}(\tau,{\bf k})
= \frac{8\pi^2G}{k^2}\int^{x}_{x_i}dy(yx)^{1/2}[N_{\nu}(x)J_{\nu}(y)-J_{\nu}(x)N_{\nu}(y)]a(y)T^{\mathrm{TT}}_{ij}(y,{\bf k}) \quad (\mathrm{for}\ \tau\le\tau_f), \label{eq2-8}
\end{align}
\end{widetext}
where $J_{\nu}(x)$ and $N_{\nu}(x)$ are Bessel function and Neumann function, respectively. 
Note that this is just a generalization of the Green function solution obtained in \cite{2007PhRvD..76l3517D} for the radiation dominated background.
Substituting $\nu=1/2$, one can easily check that eq.~(\ref{eq2-8}) reduces to the result in the radiation dominated universe derived in \cite{2007PhRvD..76l3517D}.

After the time $\tau_f$, the source term becomes negligible in eq.~(\ref{eq2-6}), and $\bar{h}_{ij}$ is given by
a linear combination of two independent solutions of eq.~(\ref{eq2-6}) without the source term
\begin{align}
&\bar{h}_{ij}(\tau,{\bf k})
= A_{ij}({\bf k})(k\tau)^{1/2}J_{\nu}(k\tau)\notag\\
& \quad\qquad\qquad+ B_{ij}({\bf k})(k\tau)^{1/2}N_{\nu}(k\tau) \quad (\mathrm{for}\ \tau\ge\tau_f). \label{eq2-9}
\end{align}
The coefficients $A_{ij}$ and $B_{ij}$ are determined by matching the solution given by eq.~(\ref{eq2-8}) with eq.~(\ref{eq2-9})
at $\tau=\tau_f$. We obtain
\begin{align}
A_{ij}({\bf k}) &= -\frac{8\pi^2G}{k^2}\int^{x_f}_{x_i}dx\sqrt{x}a(x)N_{\nu}(x)T^{\mathrm{TT}}_{ij}(x,{\bf k}), \notag\\
B_{ij}({\bf k}) &= \frac{8\pi^2G}{k^2}\int^{x_f}_{x_i}dx\sqrt{x}a(x)J_{\nu}(x)T^{\mathrm{TT}}_{ij}(x,{\bf k}). \label{eq2-10}
\end{align}

Let us define a dimensionless anisotropic stress tensor
\begin{equation}
a^{-2}(\tau)T_{ij}^{\mathrm{TT}}(\tau,{\bf k}) = (\rho+p)\Pi_{ij}(\tau,{\bf k}), \label{eq2-11}
\end{equation}
where $\rho$ is the background homogeneous energy density and $p$ is the background homogeneous pressure.
We assume that the source is statistically homogeneous and isotropic, and introduce the unequal time
correlator of the anisotropic stress tensor
\begin{align}
\langle\Pi_{ij}(\tau_1,{\bf k})\Pi^*_{ij}(\tau_2,{\bf k'})\rangle \equiv (2\pi)^3\delta^{(3)}({\bf k-k}')\Pi(k,\tau_1,\tau_2), \label{eq2-12}
\end{align}
where $\langle\cdots\rangle$ denotes an ensemble average for a stochastic background.
We note the following relations
\begin{align}
\tau Ha(\tau) = \alpha = \frac{\beta}{1-\beta}, \label{eq2-13}\\
\beta = \frac{2}{3(1+w)}, \label{eq2-14}
\end{align}
where $w$ is a mean equation of state defined by $p=w\rho$.
By using these relations, we can rewrite eq.~(\ref{eq2-10}) as
\begin{align}
A_{ij}({\bf k}) &= -\frac{2\pi \beta}{(1-\beta)^2}\int^{x_f}_{x_i}dxx^{-3/2}a(x)N_{\nu}(x)\Pi_{ij}(x,{\bf k}), \notag\\
B_{ij}({\bf k}) &= \frac{2\pi \beta}{(1-\beta)^2}\int^{x_f}_{x_i}dxx^{-3/2}a(x)J_{\nu}(x)\Pi_{ij}(x,{\bf k}). \label{eq2-15}
\end{align}

The energy density of GWs is given by (see e.g. \cite{Maggiore2008})
\begin{align}
\rho_{\mathrm{gw}} &= \frac{1}{32\pi G}\langle\dot{h}_{ij}(t,{\bf x})\dot{h}_{ij}(t,{\bf x})\rangle \notag\\
&\simeq \frac{1}{32\pi Ga^4(\tau)}\langle\bar{h}'_{ij}(\tau,{\bf x})\bar{h}'_{ij}(\tau,{\bf x})\rangle, \label{eq2-16}
\end{align}
where we neglected the terms with higher order in $aH$ in the second equality, since we 
assume that the wavelength of GWs is well inside the Hubble radius at the time $\tau$ ($k\tau\gg 1$).
Substituting eqs.~(\ref{eq2-9}) and (\ref{eq2-15}) into eq.~(\ref{eq2-16}), and using eq.~(\ref{eq2-12}), we obtain
\begin{widetext}
\begin{eqnarray}
\rho_{\mathrm{gw}}(t)
&=& \frac{1}{32\pi Ga^4(t)}\int\frac{d^3{\bf k}}{(2\pi)^3}\frac{k^2}{\pi}\frac{4\pi^2\beta^2}{(1-\beta)^4} \notag\\
&&\times \left\{ \int^{x_f}_{x_i}\frac{dx_1}{x^{3/2}_1}a(x_1)N_{\nu}(x_1)\int^{x_f}_{x_i}\frac{dx_2}{x^{3/2}_2}a(x_2)N_{\nu}(x_2)\Pi(k,\tau_1,\tau_2) \right.\notag\\
&&\qquad \left.+  \int^{x_f}_{x_i}\frac{dx_1}{x^{3/2}_1}a(x_1)J_{\nu}(x_1)\int^{x_f}_{x_i}\frac{dx_2}{x^{3/2}_2}a(x_2)J_{\nu}(x_2)\Pi(k,\tau_1,\tau_2)\right\}, \label{eq2-17}
\end{eqnarray}
where we used the approximations for $k\tau\gg1$
\begin{equation}
J_{\nu}(k\tau) \to \sqrt{\frac{2}{\pi(k\tau)}}\cos\left(k\tau - \frac{\nu\pi}{2} - \frac{\pi}{4}\right), \quad
N_{\nu}(k\tau) \to \sqrt{\frac{2}{\pi(k\tau)}}\sin\left(k\tau - \frac{\nu\pi}{2} - \frac{\pi}{4}\right), \label{eq2-18}
\end{equation}
\end{widetext}
and averaged over a period of the oscillation of sine and cosine with time.
We define the fraction of the energy density of GWs at the time $t$ as
\begin{equation}
\Omega_{\mathrm{gw}}(t) = \frac{1}{\rho(t)}\frac{d\rho_{\mathrm{gw}}(t)}{d\ln k}. \label{eq2-19}
\end{equation}
Note that, it is not the value which would be observed at the present time.
We introduce this notation for convenience to present the result of numerical simulations.
We will convert it into the spectrum of GWs observed today in section~\ref{sec4-4}.
Substituting eq.~(\ref{eq2-17}) into eq.~(\ref{eq2-19}), we finally obtain
\begin{widetext}
\begin{eqnarray}
\Omega_{\mathrm{gw}}(t) &=& \frac{k^3}{6\pi(1-\beta)^2}\int^{x_f}_{x_i}\frac{dx_1}{\sqrt{x_1}}\frac{x}{x_1}\frac{a(x_1)}{a(x)}\int^{x_f}_{x_i}\frac{dx_2}{\sqrt{x_2}}\frac{x}{x_2}\frac{a(x_2)}{a(x)}\left[N_{\nu}(x_1)N_{\nu}(x_2)+ J_{\nu}(x_1)J_{\nu}(x_2)\right]\Pi(k,\tau_1,\tau_2), \label{eq2-20}
\end{eqnarray}
\end{widetext}
where we used $\rho(t)=\frac{3}{8\pi G}H^2(t)$.

In the previous work~\cite{2010JCAP...05..032H}, the spectrum of GWs was directly obtained by calculating TT pert of the stress-energy tensor $T^{\mathrm{TT}}_{ij}$
in numerical simulations. Another way to evaluate the spectrum of GWs is to estimate the anisotropic stress power spectrum $\Pi(k,\tau_1,\tau_2)$
and use eq.~(\ref{eq2-20}). We will evaluate in both ways and compare the results.
The advantage of using eq.~(\ref{eq2-20}) is that we can decompose the origin of $k$ dependence of $\Omega_{\mathrm{gw}}$:
The tilt of the spectrum is determined by the $k$ dependence in $\Pi(k,\tau_1,\tau_2)$ and the time integral of the Green function.
This decomposition might be helpful to understand the precise form of the GW spectrum, as we see in section~\ref{sec4}.

%%%%%%%%%%%%%%%%%%%%%%%%%%%%%%%%%%%%%%%%%%%%%%%%%%%%%%%%%%%%%%%%%%%%%%
\section{\label{sec3}Numerical simulation of domain walls}
%%%%%%%%%%%%%%%%%%%%%%%%%%%%%%%%%%%%%%%%%%%%%%%%%%%%%%%%%%%%%%%%%%%%%%
In this section, we show the results of numerical simulations.
We consider the simple model of real scalar field $\phi$ in which a discrete $Z_2$ symmetry is spontaneously broken.
The evolution of $\phi$ in the expanding background is described by the Klein-Gordon equation
\begin{equation}
\ddot{\phi}+3H\dot{\phi}-\frac{\nabla^2}{a^2}\phi + \frac{dV}{d\phi} = 0, \label{eq3-1}
\end{equation}
where the potential is given by
\begin{equation}
V(\phi) = \frac{\lambda}{4}(\phi^2-\eta^2)^2. \label{eq3-2}
\end{equation}
This potential has $Z_2$ symmetry under which the scalar field transforms as $\phi\to-\phi$.
This symmetry is spontaneously broken and scalar field gets vacuum expectation value $\langle\phi\rangle=\pm\eta$.
Domain walls can be formed around the region where the value of the classical field changes from $-\eta$ to $+\eta$.

If such domain walls were created in the early universe and survived until today, they eventually come to
overclose the energy density of the universe and disturb the success of standard cosmology~\cite{1974JETP...40....1Z}.
One way to avoid this problem is to introduce a term in the potential which explicitly breaks the discrete symmetry
and lifts the degeneracy of vacua~\cite{1981PhRvD..23..852V,1989PhRvD..39.1558G}.
If such a term exists, walls become unstable and eventually disappear.
We model this effect by adding a term
\begin{equation}
\delta V = \epsilon\eta\phi\left(\frac{1}{3}\phi^2-\eta^2\right), \label{eq3-3}
\end{equation}
to the potential given by eq.~(\ref{eq3-2}). The dimensionless parameter $\epsilon$, which we call ``bias",
controls the magnitude of energy difference between two vacua and determines the life time of domain walls.
We assume that $\epsilon$ is much smaller than $1$ since we are interested in the circumstance in which 
the discrete symmetry is held approximately. In particular, the condition $\epsilon<0.15\lambda$ must be satisfied
in order that the infinite size of domain is formed~\cite{2010JCAP...05..032H}.

%%%%%%%%%%%%%%%%%%%%%%%%%%%%%%%%%%%%%%%%%%%%%%%%%%%%%%%%%%%%%%%%%%%%%%
\subsection{\label{sec3-1}Initial conditions}
%%%%%%%%%%%%%%%%%%%%%%%%%%%%%%%%%%%%%%%%%%%%%%%%%%%%%%%%%%%%%%%%%%%%%%
Eq.~(\ref{eq3-1}), which describes the evolution of domain walls, is highly nonlinear
and difficult to solve analytically. Then we solve eq.~(\ref{eq3-1}) numerically on the three dimensional lattice.
In the previous study~\cite{2010JCAP...05..032H}, scalar field is assumed to be in thermal equilibrium with temperature $T$,
and the initial field configurations
are generated by considering finite temperature effects. 
In this primary thermal stage, the scalar field fluctuations also produce GWs
with the spectrum peaked at the frequency corresponding to the mass of the scalar field $\sim\sqrt{\lambda}\eta$.  
However, this setup might be irrelevant to our interest
to calculate GW spectrum produced by domain walls. The reason is as follows:
The amplitude of GWs becomes large enough to observe if domain walls survived for sufficiently long time. This means that
the spectrum of GWs produced at the primary stage is negligible compared with that produced by domain walls at the late time.
Furthermore, the assumption that the phase transition occurred at $T\sim\eta$ gives a severe constraint on the range of parameters
which we choose to perform realistic simulations. In numerical simulations, we must resolve the width of the wall $\delta_w\sim\eta^{-1}$
and keep Hubble radius $H^{-1}$ smaller than the size of the simulation box. If we assume that the temperature is given by $T\sim\eta$,
the ratio of these two length scale is $\delta_w/H^{-1}\sim\eta/M_P$, where $M_P$ is the Planck mass.
Therefore, we must choose $\eta$ close to the Planck scale in order to maintain the resolution of the width of domain walls.
In numerical simulations performed in~\cite{2010JCAP...05..032H}, the value of $\eta$ is chosen to be $10^{17}$GeV,
and the results of the simulations are extrapolated into lower values of $\eta$.
However, it is not obvious that this extrapolation is held for arbitrary scale of $\eta$.

To avoid these difficulties, we omit the assumption of thermal initial conditions and give the initial field configurations 
as Gaussian random amplitudes. In this case, we expect that only domain walls contribute as a source of GWs.
In addition, if we normalize all the dimensionful quantities in the unit of $\eta$, the results of numerical simulations
become independent of $\eta$. See appendix \ref{secA-1} for more details of the setup of the simulations.

%%%%%%%%%%%%%%%%%%%%%%%%%%%%%%%%%%%%%%%%%%%%%%%%%%%%%%%%%%%%%%%%%%%%%%
\subsection{\label{sec3-2}Evolution of the domain wall networks}
%%%%%%%%%%%%%%%%%%%%%%%%%%%%%%%%%%%%%%%%%%%%%%%%%%%%%%%%%%%%%%%%%%%%%%
Now, we present the results of numerical simulations. 
We performed lattice simulations with $256^3$ points in both radiation and matter dominated backgrounds.
The comoving size of the simulation box is set to be 50 (in the radiation dominated era) or 24 (in the matter dominated era)
in the unit of $\eta^{-1}$. We fix $\lambda=0.1$ and vary the value of $\epsilon$. 
The initial time (in the unit of $\eta^{-1}$) is set to be $t_i=1$.
The final time is set to be $t_f=151$ in the simulation with radiation dominated background and $t_f=65$ in the simulation
with matter dominated background.

\begin{figure*}[htp]
\centering
$\begin{array}{cc}
\subfigure[]{
\includegraphics[width=0.45\textwidth]{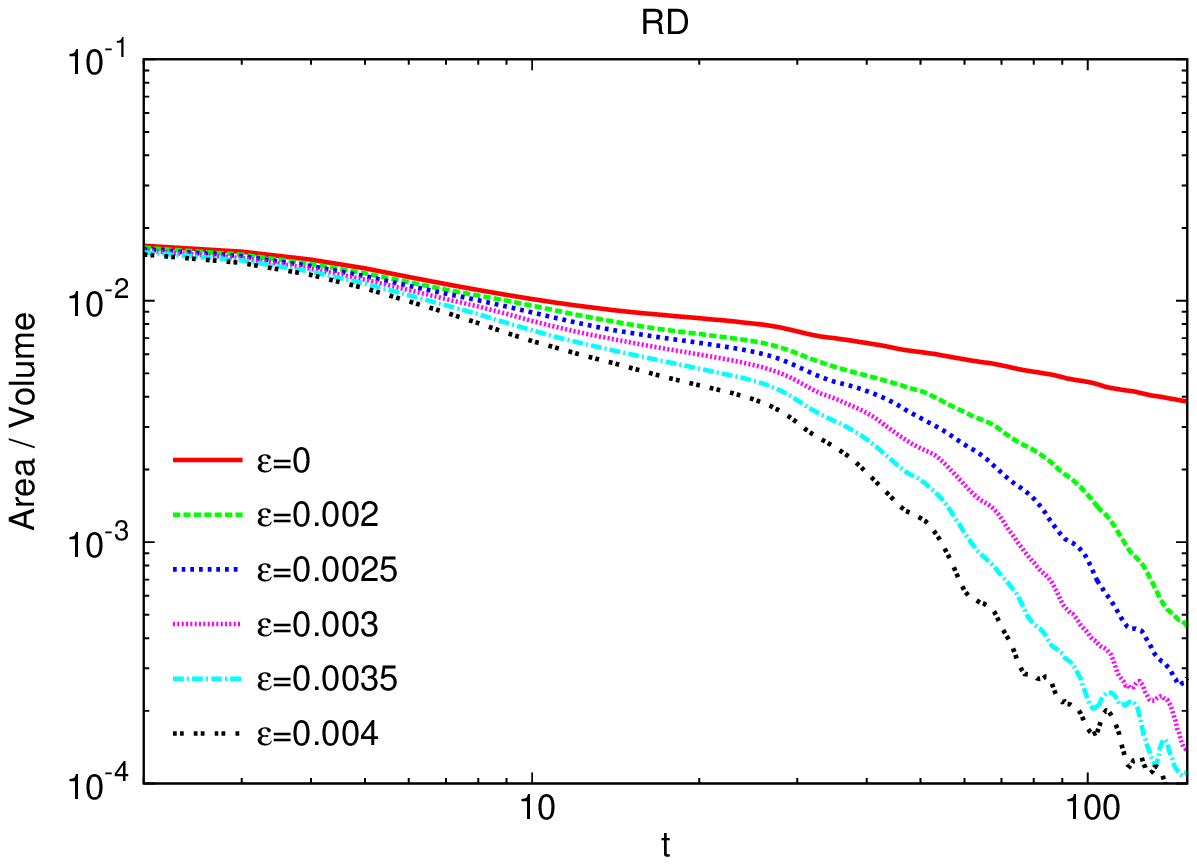}}
\hspace{20pt}
\subfigure[]{
\includegraphics[width=0.45\textwidth]{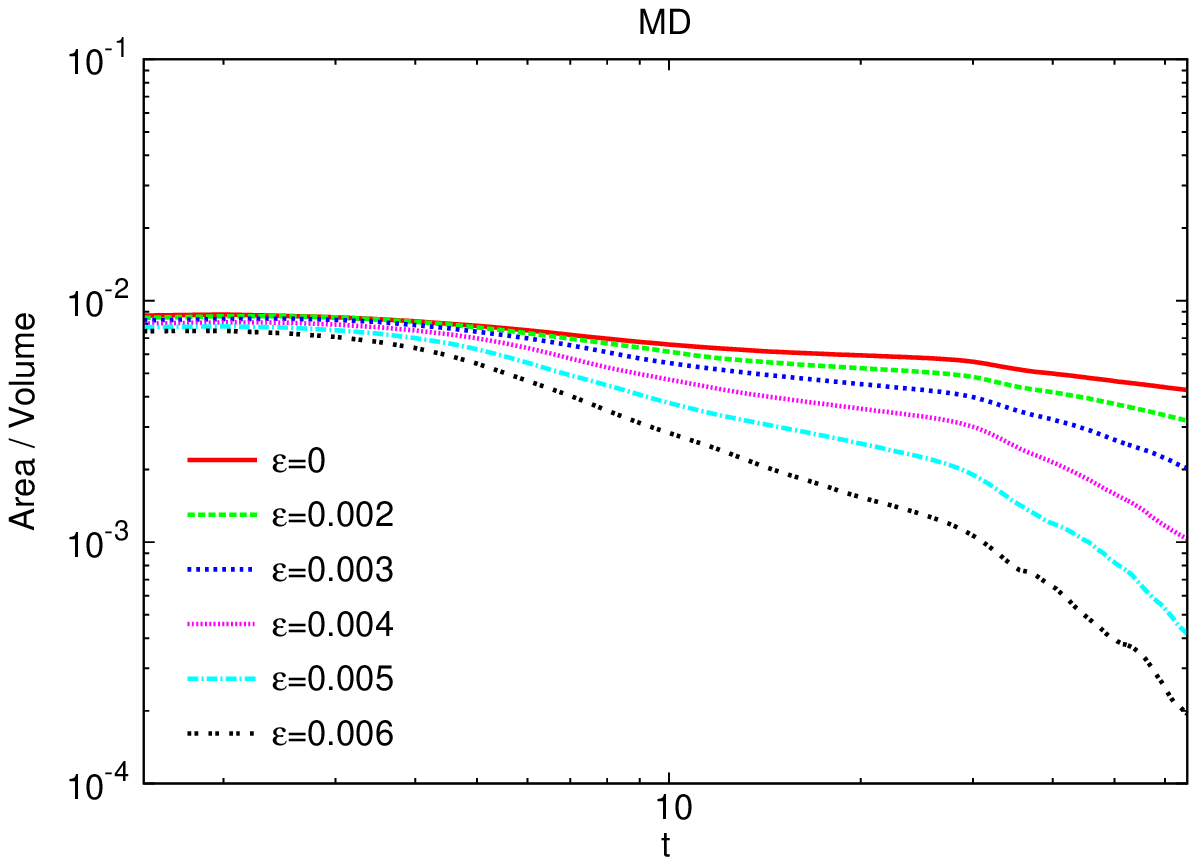}}
\end{array}$
\caption{The time evolution of the comoving area density of domain walls for various values of $\epsilon$ in (a) radiation dominated background and (b) matter dominated background.}
\label{fig1}
\end{figure*}

\begin{figure*}[htp]
\centering
$\begin{array}{cc}
\subfigure[]{
\includegraphics[width=0.45\textwidth]{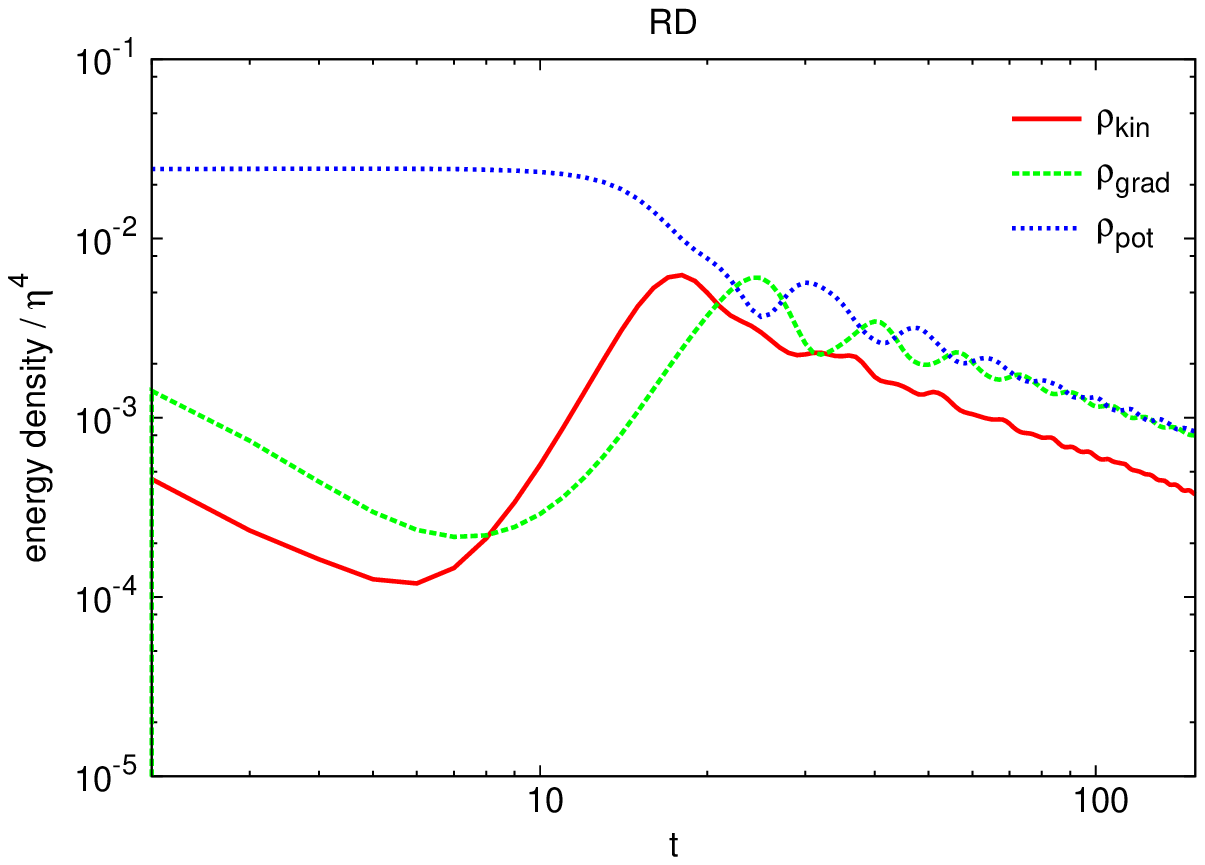}}
\hspace{20pt}
\subfigure[]{
\includegraphics[width=0.45\textwidth]{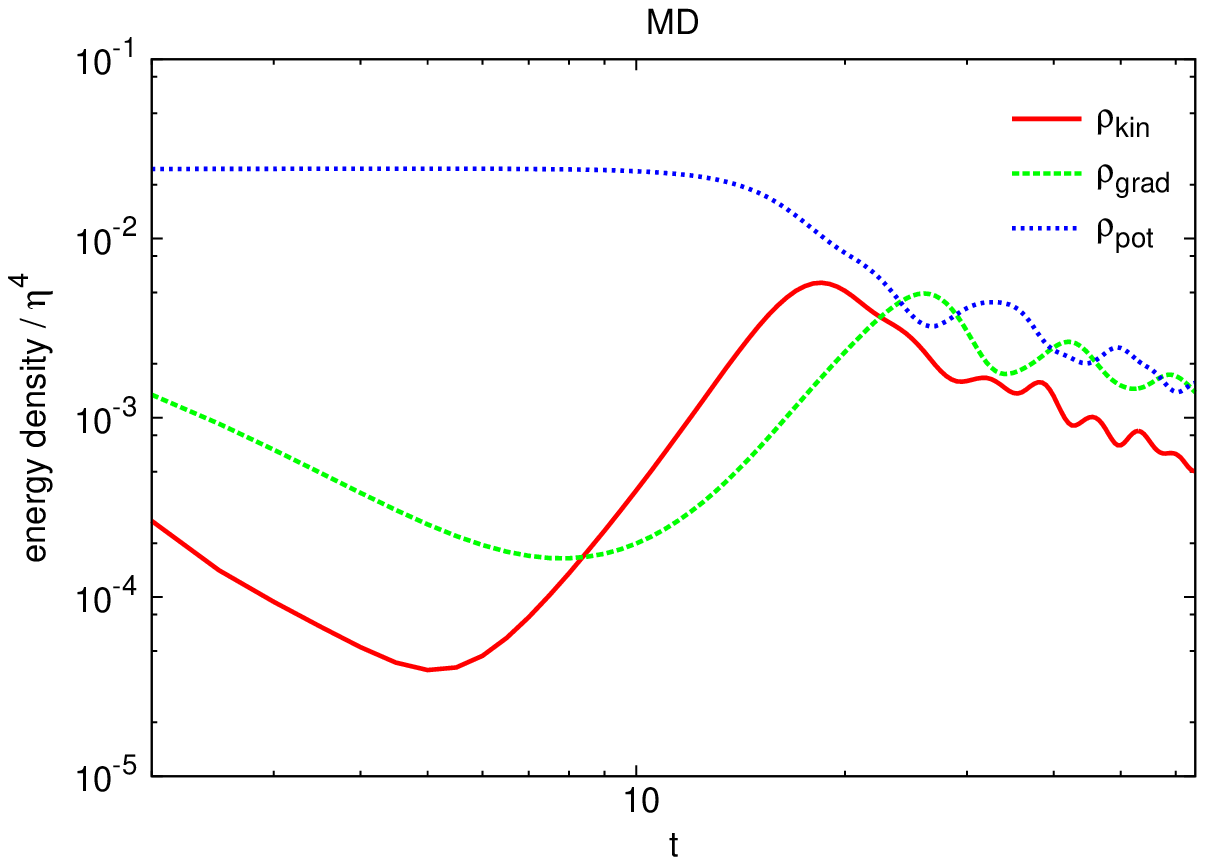}}
\end{array}$
\caption{The time evolution of the kinetic/gradient/potential energy densities of domain walls (in the unit of $\eta^4$) in (a) radiation dominated background and (b) matter dominated background.}
\label{fig2}
\end{figure*}

Figure~\ref{fig1} shows the time evolution of the area occupied by domain walls in the simulation box.
We see that, if $\epsilon\ne0$, the area density of domain walls decays at late time. 
It means that domain walls collapse due to the existence of the bias.
On the other hand, if $\epsilon=0$, the comoving area density of domain walls evolves as $\propto\tau^{-1}$.
This property is called the scaling solution~\cite{1989ApJ...347..590P,1996PhRvL..77.4495H,2003PhRvD..68j3506G,2005PhLB..610....1A,2005PhRvD..71h3509O} and corresponds to the fact that the energy density of domain walls evolves like
$\rho_{\mathrm{wall}}\sim\sigma/t$, where $\sigma=2\sqrt{2\lambda}\eta^3/3$ is the surface mass density of the domain wall.
Figure~\ref{fig2} shows the time evolution of the energy density of the scalar field.
From this figure, we see that the scaling regime in which energy densities evolve as $\propto 1/t$ begins at around $t\simeq 20$.

Note that, in our study, the dynamical range of the simulation is quite short.
This is due to the fact that we can not resolve the width of domain walls for a long time in the
comoving box if we take account of the cosmic expansion.
In particular, the actual dynamical range in the simulation with matter dominated background is as small as
$(t_f/t_{\mathrm{form}})^{1/2}\simeq(65/20)^{1/2}\simeq1.8$ in conformal time, where $t_{\mathrm{form}}$
is the time when domain walls enter the scaling regime.
Future simulations with higher spatial resolutions should
improve the dynamical range and confirm our current results of numerical simulations.

\subsection{\label{sec3-3}Spectrum of gravitational waves}
We calculate the spectrum of GWs directly from numerical simulations.
The method of the calculation is summarized in Appendix~\ref{secA-2}.
Here we present the results of the direct calculations and briefly discuss the features of the spectrum of GWs.
We will reevaluate the spectrum of GWs in the next section by using another formalism.

\begin{figure*}[htp]
\centering
$\begin{array}{cc}
\subfigure[]{
\includegraphics[width=0.45\textwidth]{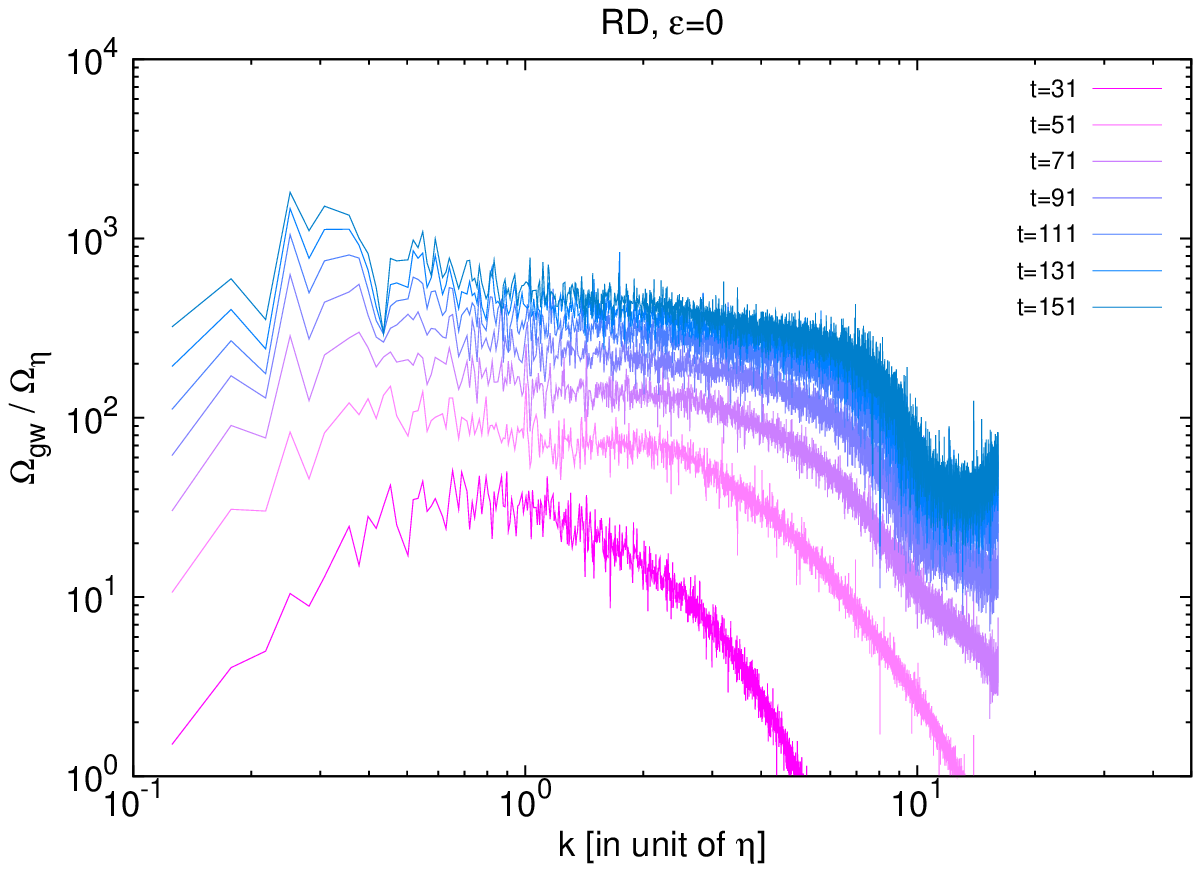}}
\hspace{20pt}
\subfigure[]{
\includegraphics[width=0.45\textwidth]{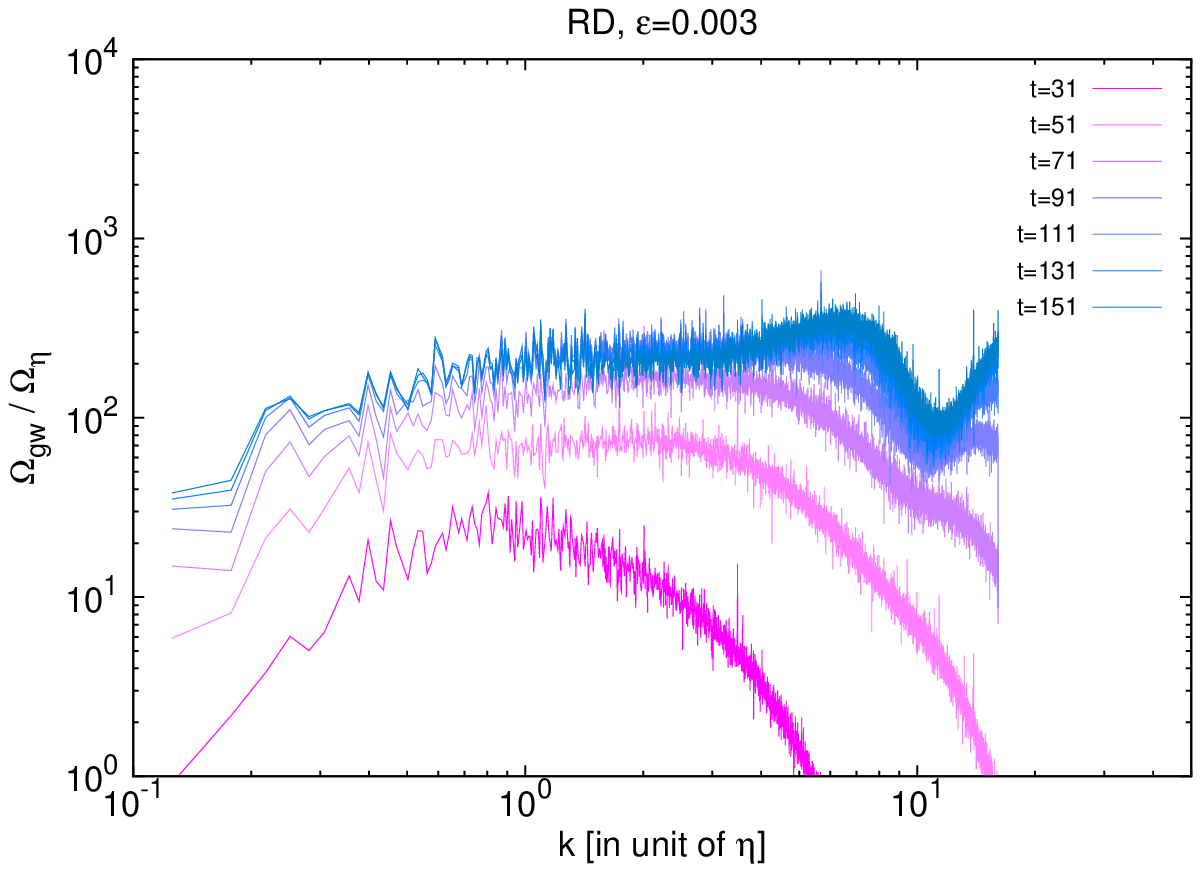}} \\
\subfigure[]{
\includegraphics[width=0.45\textwidth]{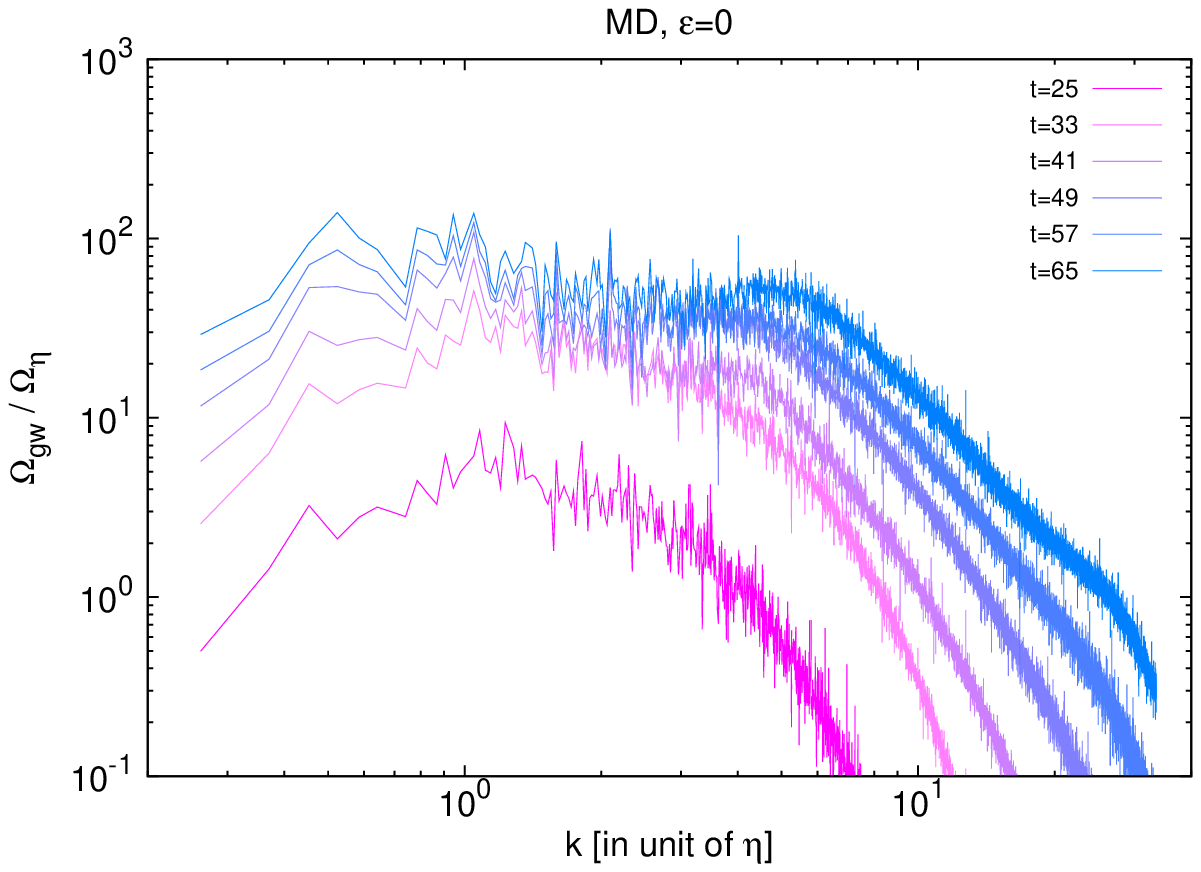}}
\hspace{20pt}
\subfigure[]{
\includegraphics[width=0.45\textwidth]{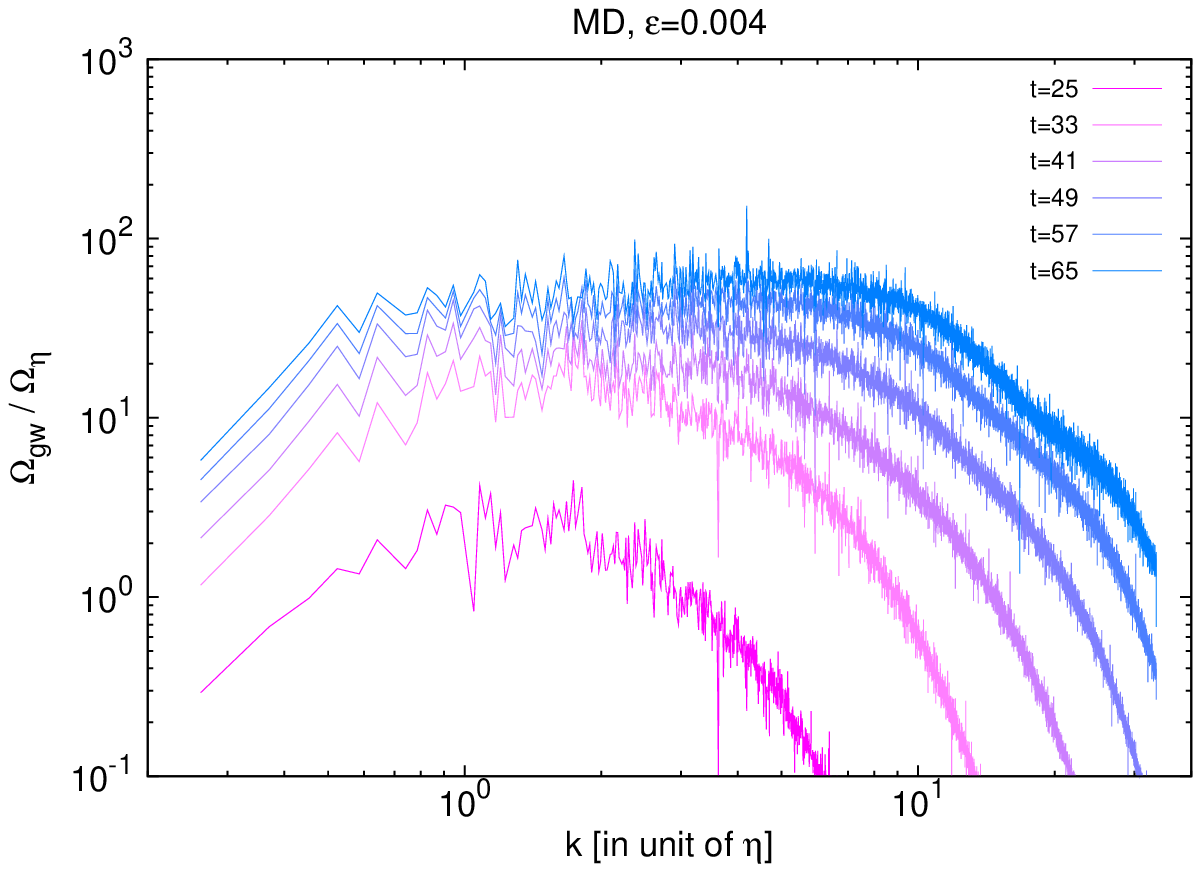}}
\end{array}$
\caption{The spectra of gravitational waves obtained by numerical simulations. The top panel shows the results in radiation dominated background
with (a) $\epsilon=0$ and (b) $\epsilon=0.003$, and the bottom panel shows that in matter dominated background with (c) $\epsilon=0$ and (d) $\epsilon=0.004$. The different colors correspond to the spectra at different time. The spectra are shown from the time $t=31$ (pink) to $t=151$ (green)
with the interval $\Delta t=20$ in the simulation with radiation dominated background, and from the time $t=25$ (pink) to $t=65$ (green)
with the interval $\Delta t=8$ in the simulation with matter dominated background.}
\label{fig3}
\end{figure*}

The spectra of GWs obtained from simulations are shown in figure~\ref{fig3}.
The vertical axis represents the amplitude of GWs defined by eq.~(\ref{eq2-19}). We normalize it by the dimensionless quantity
\begin{equation}
\Omega_{\eta} \equiv \frac{\rho^{\eta}_{\mathrm{gw}}}{\rho(t_i)} = \frac{8\pi}{3\beta^2}G^2\eta^4, \label{eq3-4}
\end{equation}
where $\rho^{\eta}_{\mathrm{gw}}=G\eta^6$ is the energy density of GWs (estimated by using the quadrupole formula of GWs)
radiated by a source which has a characteristic scale $\eta$,
and $\rho(t_i)=3H^2(t_i)/8\pi G=3\beta^2\eta^2/8\pi G$ is the background homogeneous energy density at the initial time of the simulations.
We introduce this notation for a convenience to present the numerical results: $\Omega_{\mathrm{gw}}(t)/\Omega_{\eta}$ becomes ${\cal O}(1)$
at the beginning of the simulation.
The horizontal axis represents the comoving wavenumber $k$ normalized by $\eta$. This is not the frequency of GWs, which is given by $f=k/2\pi a(t)$.

From figure~\ref{fig3}, we see that the spectrum of GWs is almost flat in both radiation dominated and matter dominated backgrounds.
In the previous study~\cite{2010JCAP...05..032H}, it is conjectured that this flat spectrum extends roughly between the frequency
corresponding to the Hubble radius at the time of the decay of domain walls and that corresponding to the width of domain walls.
Let us define comoving wavenumbers $k_h$ and $k_w$ which correspond to the Hubble radius and the width of domain walls, respectively,
\begin{equation}
\frac{k_h}{a(t)} = 2\pi H(t)\quad \mathrm{and} \quad \frac{k_w}{a(t)}=2\pi\sqrt{\lambda}\eta. \label{eq3-5}
\end{equation}
Note that the values of $k_h$ and $k_w$ vary with time. 
In the present numerical simulations where all of the dimensionful quantities are normalized in the unit of $\eta$,
$k_h$ changes from $\pi$ at the initial time to $0.08\pi$ at the final time in the simulation with radiation dominated background,
and it changes from $1.3\pi$ to $0.33\pi$ in the simulation with matter dominated background.
Also, $k_w$ changes from $0.63\pi$ to $7.8\pi$ in the simulation with radiation dominated background,
and it changes from $0.63\pi$ to $10\pi$ in the simulation with matter dominated background.
Therefore, these two scales become separate at the late time of simulations.
In fact, we see that the band of the spectra shown in figure~\ref{fig3} becomes wider as time passes.

In our previous study~\cite{2010JCAP...05..032H}, the spectrum of GWs had a peak at the high frequency (around $k\lesssim k_w$),
and the slope of the spectrum was mildly blue tilted.
However, in the present results, the spectrum with $\epsilon=0$ becomes slightly red tilted as we see in figures~\ref{fig3} (a) and (c).
This difference might be caused by the fact that we use different initial conditions for numerical simulations:
The spectrum obtained in~\cite{2010JCAP...05..032H} contains the GWs produced during the primary thermal stage (see section~\ref{sec3-1}),
which is not completely negligible in the time scale of the simulations and gives a peak at the high frequency.
Therefore, we expect that if GWs from primary thermal stage become negligible and the source of GWs is purely domain walls,
the spectrum of GWs is almost flat and slightly red tilted.
Then, if domain walls collapse, the amplitude of GWs grows at the high frequency as we see in figures~\ref{fig3} (b) and (d).
This corresponds the fact that the false vacuum regions become small when domain walls decay. 

To summarize, the spectrum of GWs obtained by direct numerical calculations indicates that the shape of the spectrum
is nearly flat (or slightly red tilted) and the slope of the spectrum seems to change at two characteristic frequencies corresponding
to the Hubble radius at the decay of domain walls and the width of domain walls. 
High frequency modes grow at the late time due to the collapse of domain walls.
These properties are seen in both radiation dominated and matter dominated backgrounds.

%%%%%%%%%%%%%%%%%%%%%%%%%%%%%%%%%%%%%%%%%%%%%%%%%%%%%%%%%%%%%%%%%%%%%%
\section{\label{sec4}The spectrum of gravitational waves from domain walls: Another approach}
%%%%%%%%%%%%%%%%%%%%%%%%%%%%%%%%%%%%%%%%%%%%%%%%%%%%%%%%%%%%%%%%%%%%%%
The lattice simulations revealed that some properties of GWs produced by domain walls, such as the nearly flat spectrum
and the existence of two relevant scales (the Hubble radius at the decay of domain walls and the width of domain walls).
However, these results have many numerical uncertainties
especially in the high and low frequency bands of the spectrum, because of the lack of the spatial resolution in the lattice simulation.
In order to remove this difficulty, we take another approach to evaluate the spectrum of GWs and discuss further
about the shape of the spectrum of GWs.

In section~\ref{sec2}, we derived eq.~(\ref{eq2-20}), which indicates that we can evaluate the GW spectrum if we know the
anisotropic stress power spectrum $\Pi(k,\tau_1,\tau_2)$. It is difficult to evaluate {\it unequal time} power spectrum $\Pi(k,\tau_1,\tau_2)$
without relying some ansatzs about the unequal time correlation of the source of GWs. In section~\ref{sec4-1}, we describe
approximations to evaluate the unequal time correlation functions, which is used to estimate the GW spectrum from first order phase transitions
in the literature. Using these approximations, we reevaluate the GW spectra and compare them with the results of direct numerical calculations
in section~\ref{sec4-2}. Then, we comment on the property of the GW spectrum obtained from these analysises in section~\ref{sec4-3}.
Finally, in section~\ref{sec4-4}, we convert the results into the present-day observables.

%%%%%%%%%%%%%%%%%%%%%%%%%%%%%%%%%%%%%%%%%%%%%%%%%%%%%%%%%%%%%%%%%%%%%%
\subsection{\label{sec4-1}Approximations for the unequal time correlation functions}
%%%%%%%%%%%%%%%%%%%%%%%%%%%%%%%%%%%%%%%%%%%%%%%%%%%%%%%%%%%%%%%%%%%%%%
We would like to evaluate $\Pi(k,\tau_1,\tau_2)$ defined by eq.~(\ref{eq2-12}).
Before estimating this unequal time correlator, let us consider the equal time correlation function.
Since this quantity is given as a product of $T_{ij}^{\mathrm{TT}}(\tau,{\bf k})$, 
we hope that it also be computed directly from lattice simulations.
To be exact, we can not compute the ensemble average in the left hand side of eq.~(\ref{eq2-12})
from a single realization of the numerical simulation.
However, we approximately calculate the ensemble average by taking
average over a large volume. 
Thus, 
\begin{equation}
\int\frac{d^2\hat{k}}{4\pi}\langle\Pi_{ij}({\bf k},\tau)\Pi_{ij}^*({\bf k},\tau)\rangle \approx \int\frac{d^2\hat{k}}{4\pi}\Pi_{ij}({\bf k},\tau)\Pi_{ij}^*({\bf k},\tau) \label{eq4-1}
\end{equation}
is satisfied in the limit $V\to\infty$, where $\int d\hat{k}/4\pi$ is an average over the directions of ${\bf k}$, and $V$ is the comoving volume of the simulation box.
It is easy to show eq.~(\ref{eq4-1}), assuming Gaussian statistics (see e.g.~\cite{2008cosm.book.....Wp420}).
Using the approximation given by eq.~(\ref{eq4-1}), we obtain
\begin{equation}
\Pi(k,\tau,\tau) \approx \frac{16\pi^2G^2\beta^2}{a^4(\tau)H^4V}\int\frac{d^2\hat{k}}{4\pi}T_{ij}^{\mathrm{TT}}(\tau,{\bf k})T_{ij}^{\mathrm{TT}*}(\tau,{\bf k}), \label{eq4-2}
\end{equation}
where we replaced the factor $(2\pi)^3\delta^{(3)}(0)\simeq V$.
By using eq.~(\ref{eq4-2}), we can compute the equal time power spectrum $\Pi(k,\tau,\tau)$ directly from lattice simulations.

Next, consider the unequal time correlator $\Pi(k,\tau_1,\tau_2)$.
In the recent study of GW generation from bubble collisions~\cite{2008PhRvD..77l4015C}, it was suggested to use some approximations
for the unequal time correlator in order to evaluate the GW power spectrum. These approximations are also applied 
to evaluate GWs from turbulence and magnetic fields generated by a first-order phase
transition~\cite{2009JCAP...12..024C} (see also~\cite{2009PhRvD..79h3519C}).

There are three kinds of approximations:
\begin{enumerate}
\item {\it Totally coherent approximation}\\
The source at different time is perfectly correlated, and the unequal time correlator is given by
\begin{equation}
\Pi(k,\tau_1,\tau_2) = \sqrt{\Pi(k,\tau_1,\tau_1)}\sqrt{\Pi(k,\tau_2,\tau_2)}. \label{eq4-3}
\end{equation}
\item {\it Incoherent approximation}\\
The source at different time is not correlated, and the unequal time correlator is given by
\begin{equation}
\Pi(k,\tau_1,\tau_2) = \Pi(k,\tau_1,\tau_1)\delta(\tau_1-\tau_2)\Delta\tau, \label{eq4-4}
\end{equation}
where $\Delta\tau$ is a characteristic time scale for the duration of the source (left as a free parameter).
\item {\it Top hat approximation}\\
The source is correlated for modes with a time separation $|\tau_1-\tau_2|<x_c/k$, and the unequal time correlator is given by
\begin{align}
&\Pi(k,\tau_1,\tau_2) \notag\\
&=\Pi(k,\tau_1,\tau_1)\Theta(\tau_2-\tau_1)\Theta\left(\frac{x_c}{k}-(\tau_2-\tau_1)\right) \notag\\
&\quad +\Pi(k,\tau_2,\tau_2)\Theta(\tau_1-\tau_2)\Theta\left(\frac{x_c}{k}-(\tau_1-\tau_2)\right), \label{eq4-5}
\end{align}
where $x_c$ is a dimensionless parameter of ${\cal O}(1)$ and $\Theta(\tau)$ is the Heaviside function.
\end{enumerate}
The first two approximations are physically less motivated and just introduced for a comparison.
On the other hand, the top hat ansatz is an intermediate case between two extreme cases.
The physical interpretation of this approximation is that the correlation is lost for a time difference larger than about one wavelength.

Using these approximations, we can evaluate the unequal time correlator $\Pi(k,\tau_1,\tau_2)$
from the equal time correlator $\Pi(k,\tau,\tau)$ obtained by the numerical simulations. In the next subsection, we evaluate the GW spectrum by
combining approximations given by eqs.~(\ref{eq4-3})-(\ref{eq4-5}) and the formula for the amplitude of GWs given by eq.~(\ref{eq2-20}).

%%%%%%%%%%%%%%%%%%%%%%%%%%%%%%%%%%%%%%%%%%%%%%%%%%%%%%%%%%%%%%%%
\subsection{\label{sec4-2}Evaluation of the gravitational wave spectrum}
\begin{figure*}[htp]
\centering
$\begin{array}{cc}
\subfigure[]{
\includegraphics[width=0.45\textwidth]{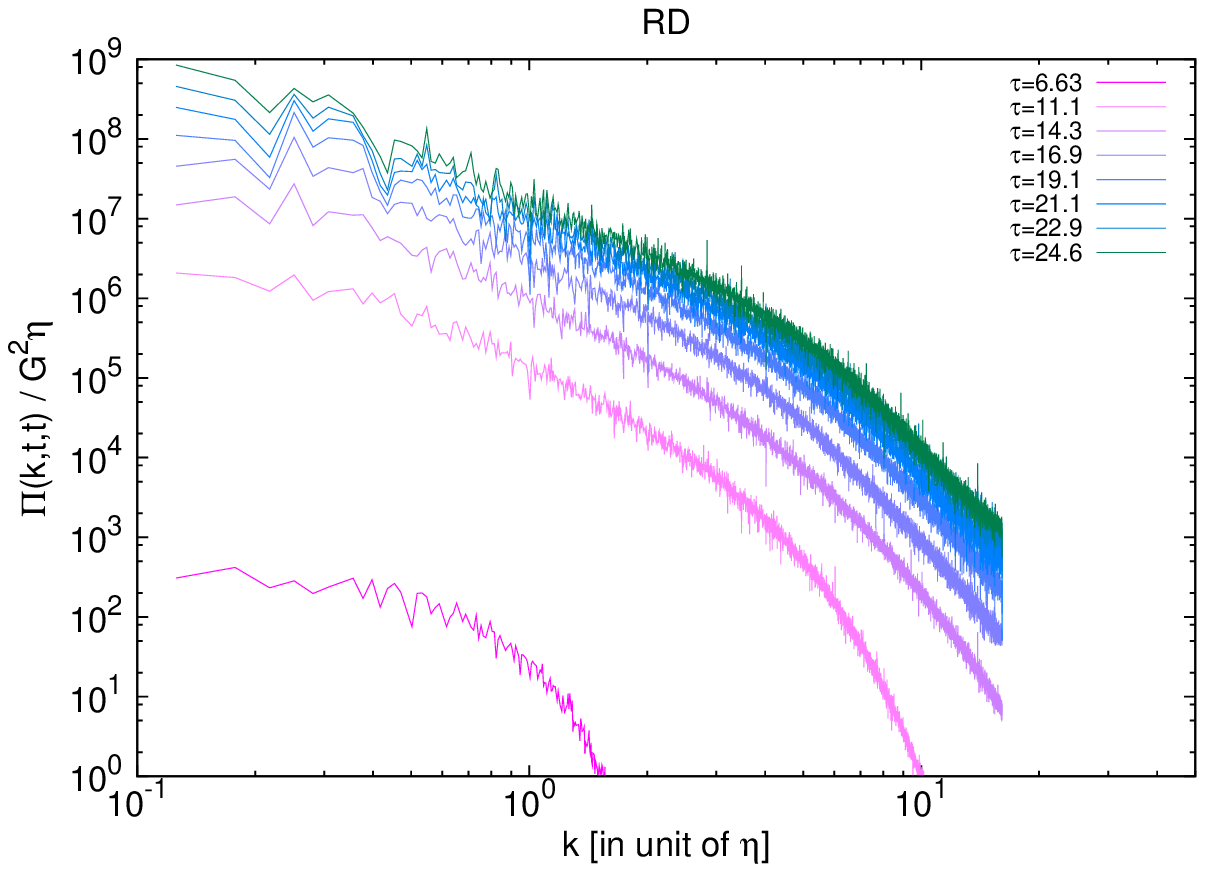}}
\hspace{20pt}
\subfigure[]{
\includegraphics[width=0.45\textwidth]{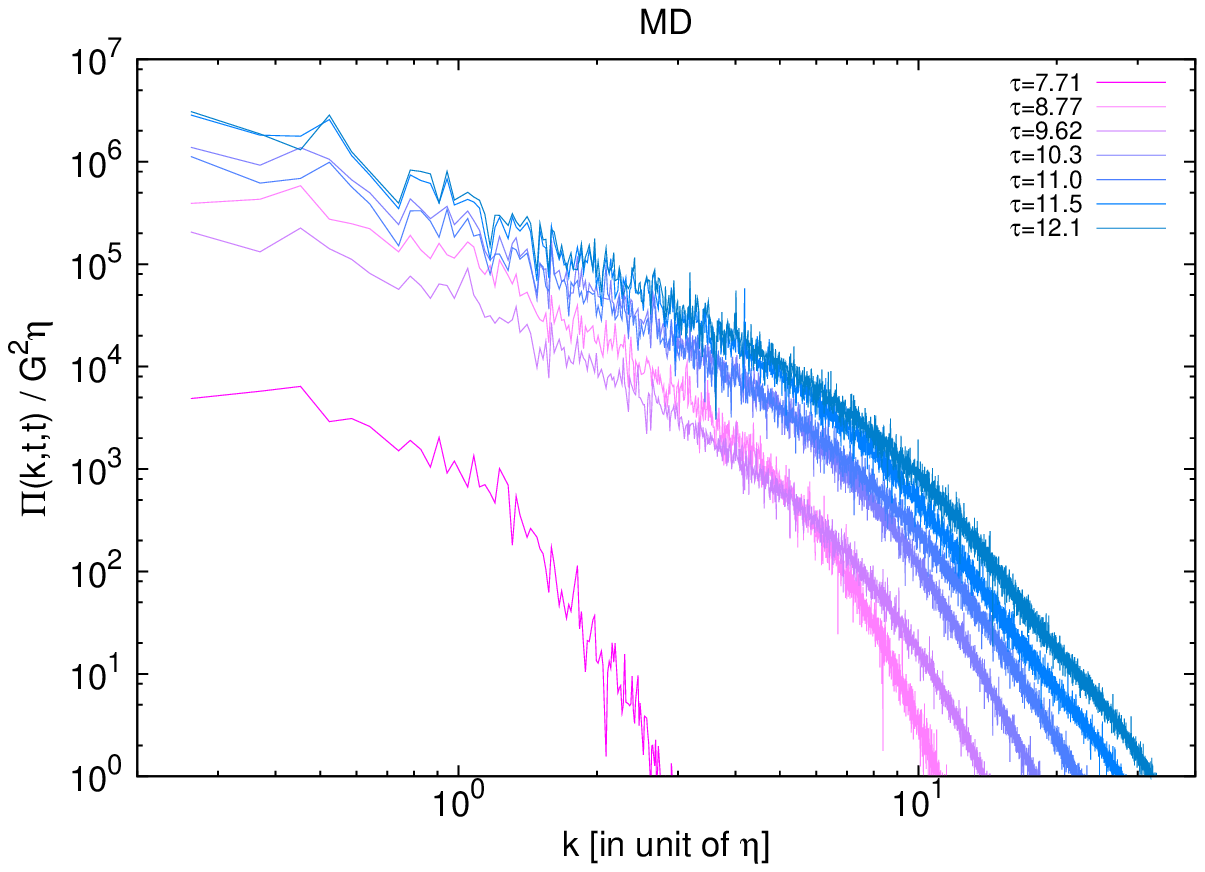}}
\end{array}$
\caption{The time evolution of the anisotropic stress power spectrum $\Pi(k,\tau,\tau)$ (divided by $G^2\eta$) obtained from numerical simulations in
(a) radiation dominated background and (b) matter dominated background.}
\label{fig4}
\end{figure*}

%%%%%%%%%%%%%%%%%%%%%%%%%%%%%%%%%%%%%%%%%%%%%%%%%%%%%%%%%%%%%%%%
First, let us evaluate the equal time anisotropic stress power spectrum $\Pi(k,\tau,\tau)$.
By using eq.~(\ref{eq4-2}), we can compute $\Pi(k,\tau,\tau)$ from lattice simulations.
The result is shown in figure~\ref{fig4}. From this figure, we see that the tilt of the spectrum becomes steeper in the small scale.
In section~\ref{sec3-3}, we argued that the spectrum of GWs is determined by two characteristic scales: the width of the wall and the Hubble radius.
Therefore, we expect that the tilt of the power spectrum $\Pi(k,\tau,\tau)$ also changes at these characteristic scales.
Furthermore, since the source has no correlation for the length scale beyond the Hubble radius,
the power spectrum would become independent of $k$ in the large scale limit
(this is just an assumption, but one can show analytically that $\Pi(k\to 0)$
is independent of $k$ for a spherical bubble configuration~\cite{2008PhRvD..77l4015C}).
Based on these considerations, we express the $k$ dependence of the power spectrum
obtained by the numerical simulation at the final time $\tau_f$ as
\begin{widetext}
\begin{align}
\Pi(k,\tau_f,\tau_f)/G^2\eta = A\left[\left(1+\left(\frac{K_h(\tau_f)}{B}\right)^{C}\right)\left(1+\left(\frac{K_w(\tau_f)}{D}\right)^{E}\right)\right]^{-1}, \label{eq4-6}
\end{align}
\end{widetext}
where $K_h(\tau_f)$ and $K_w(\tau_f)$ are given by
\begin{align}
&K_h(\tau_f) = k/k_h(\tau_f) = \left(\frac{k}{a}\right)\left/\left(\frac{2\pi}{H^{-1}}\right)\right|_{\tau=\tau_f}, \notag\\
&K_w(\tau_f) = k/k_w(\tau_f) = \left(\frac{k}{a}\right)\left/\left(\frac{2\pi}{(\sqrt{\lambda}\eta)^{-1}}\right)\right|_{\tau=\tau_f}. \label{eq4-7}
\end{align}
These are the ratio between the comoving momentum $k$ and $k_h$, or $k_w$, defined by eq.~(\ref{eq3-5}) at the conformal time $\tau_f$.
We fit the result of $\Pi(k,\tau_f,\tau_f)$ obtained from numerical simulations to the function (\ref{eq4-6}) by using the least squares method
and determine unknown parameters $A$, $B$, $C$, $D$ and $E$ (see appendix \ref{secA-3} for details). The results are shown in table~\ref{tab1}.
The parameters are fixed with uncertainties of ${\cal O}$(1) - ${\cal O}$(0.1)\%.

\setlength{\tabcolsep}{10pt}
\begin{table*}[htp]
\begin{center} 
\caption{The values of the parameters in eq.~(\ref{eq4-6}) determined by the least squares fitting. $\tilde{A}$ is the amplitude
of $\Pi(k,\tau,\tau)/G^2\eta$ at $\tau=\eta^{-1}$, as defined in eqs.~(\ref{eq4-9}) and (\ref{eq4-10}).}
\begin{tabular}{c c c}
\hline\hline
& radiation dominated & matter dominated \\
\hline\hline
$A$ & (1.84$\pm$0.09)$\times$10$^9$ & (1.57$\pm$0.03)$\times$10$^6$\\
\hline
$B$ & 0.388$\pm$0.008 & 0.724$\pm$0.004\\
\hline
$C$ & 2.020$\pm$0.005 & 2.583$\pm$0.002\\
\hline
$D$ & 0.1866$\pm$0.0007 & 0.2813$\pm$0.0008\\
\hline
$E$ & 3.1448$\pm$0.0004 & 3.5444$\pm$0.0009\\
\hline
$\tilde{A}$ & (3.4$\pm$0.2)$\times$10$^{-1}$ & (2.9$\pm$0.1)$\times$10$^{-4}$ \\
\hline\hline
\label{tab1}
\end{tabular}
\end{center}
\end{table*}

We note that the amplitude of $\Pi(k,\tau,\tau)$ grows with time, as we see in figure~\ref{fig4}.
To include this property, we model the form of the equal time anisotropic stress power spectrum as
\begin{equation}
\Pi(k,\tau,\tau) \propto \left(\frac{\rho_{\mathrm{grad}}}{\rho}\right)^2L^3S(k,\tau), \label{eq4-8}
\end{equation}
where $\rho_{\mathrm{grad}}$ is the gradient energy density of the source, $L$ is a characteristic scale of the problem,
and $S(k,\tau)$ is a dimensionless function of $k$ (and $\tau$)~\cite{2009JCAP...12..024C}.
Since the characteristic scale of domain walls is given by the Hubble radius, it is natural to expect that $L$ is as much as $\tau$.
The scaling solution also implies that $\rho_{\mathrm{grad}}$ evolves as $\propto t^{-1} \propto \tau^{-2}$,
while $\rho$ evolves as $\propto t^{-2}\propto\tau^{-4}$ in radiation dominated background. 
Similarly, we expect $\rho_{\mathrm{grad}}\propto \tau^{-3}$ and $\rho\propto\tau^{-6}$ in matter dominated background.
Therefore, we expect that the amplitude of the power spectrum evolves as
$(\rho_{\mathrm{grad}}/\rho)^2L^3\propto \tau^{\gamma}$, where $\gamma=7$ in radiation dominated background,
and $\gamma=9$ in matter dominated background.
Combining it with eq.~(\ref{eq4-6}), we obtain the following expression for the equal time correlator $\Pi(k,\tau,\tau)$
\begin{equation}
\Pi(k,\tau,\tau) = G^2\eta(\tau\eta)^{\gamma}S(k,\tau), \label{eq4-9}
\end{equation}
\begin{equation}
S(k,\tau) \simeq \tilde{A}\left[\left(1+\left(\frac{K_h(\tau)}{B}\right)^{C}\right)\left(1+\left(\frac{K_w(\tau)}{D}\right)^{E}\right)\right]^{-1}, \label{eq4-10}
\end{equation}
\begin{equation}
K_h(\tau) = \left(\frac{k}{a}\right)\left/\left(\frac{2\pi}{H^{-1}}\right)\right., \label{eq4-11}
\end{equation}
\begin{equation}
K_w(\tau) = \left(\frac{k}{a}\right)\left/\left(\frac{2\pi}{(\sqrt{\lambda}\eta)^{-1}}\right)\right., \label{eq4-12}
\end{equation}
where $\tilde{A}$ is a rescaled parameter whose value is given in table~\ref{tab1}.
Since $\tau$ is normalized by $\eta^{-1}$ in numerical simulations,
we write the $\tau$-dependent factor as $(\tau\eta)^{\gamma}$ in eq.~(\ref{eq4-9}).
We also assume that $S(k,\tau)$ depends on $\tau$ through the functions $K_h(\tau)$ and $K_w(\tau)$,
which are obtained by extrapolating the ratios (\ref{eq4-7}) into general time $\tau$.

\begin{figure*}[htp]
\centering
$\begin{array}{cc}
\subfigure[]{
\includegraphics[width=0.45\textwidth]{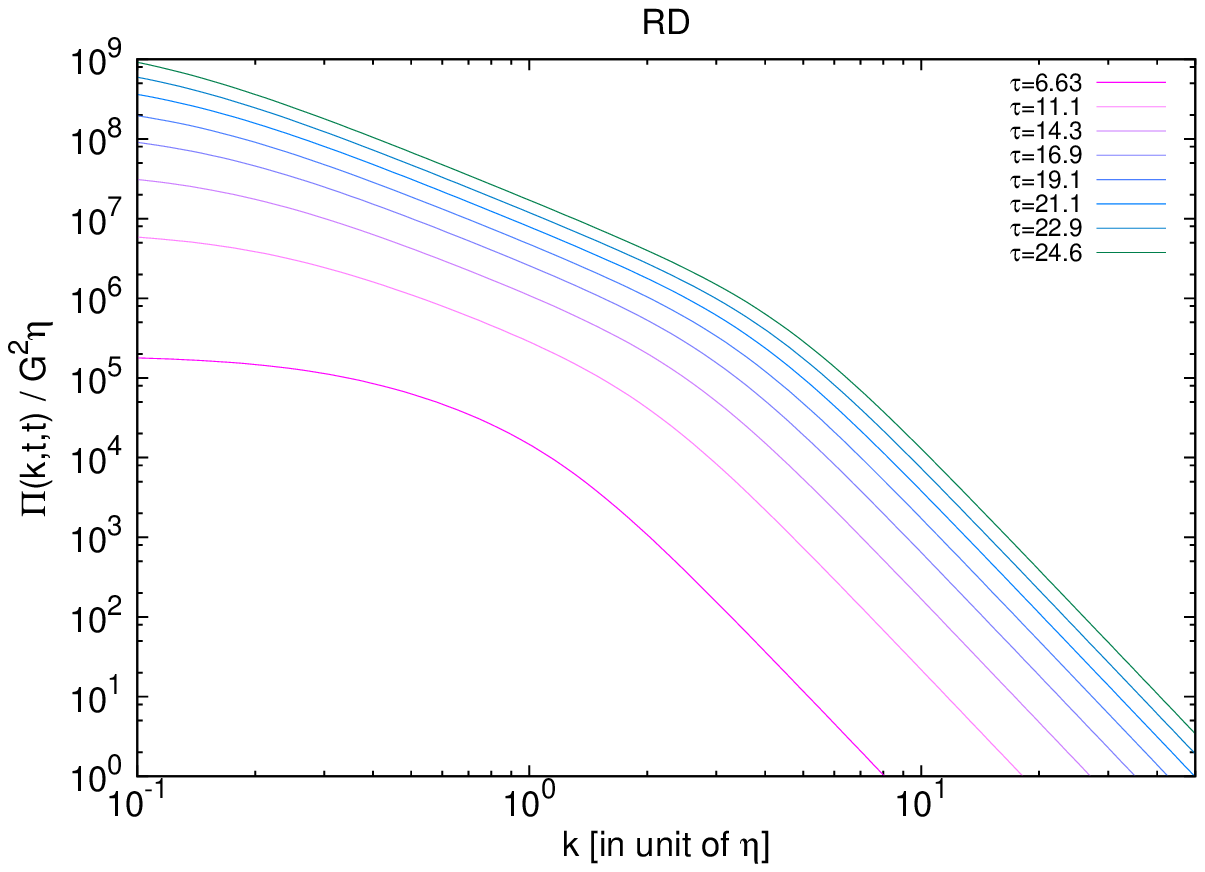}}
\hspace{20pt}
\subfigure[]{
\includegraphics[width=0.45\textwidth]{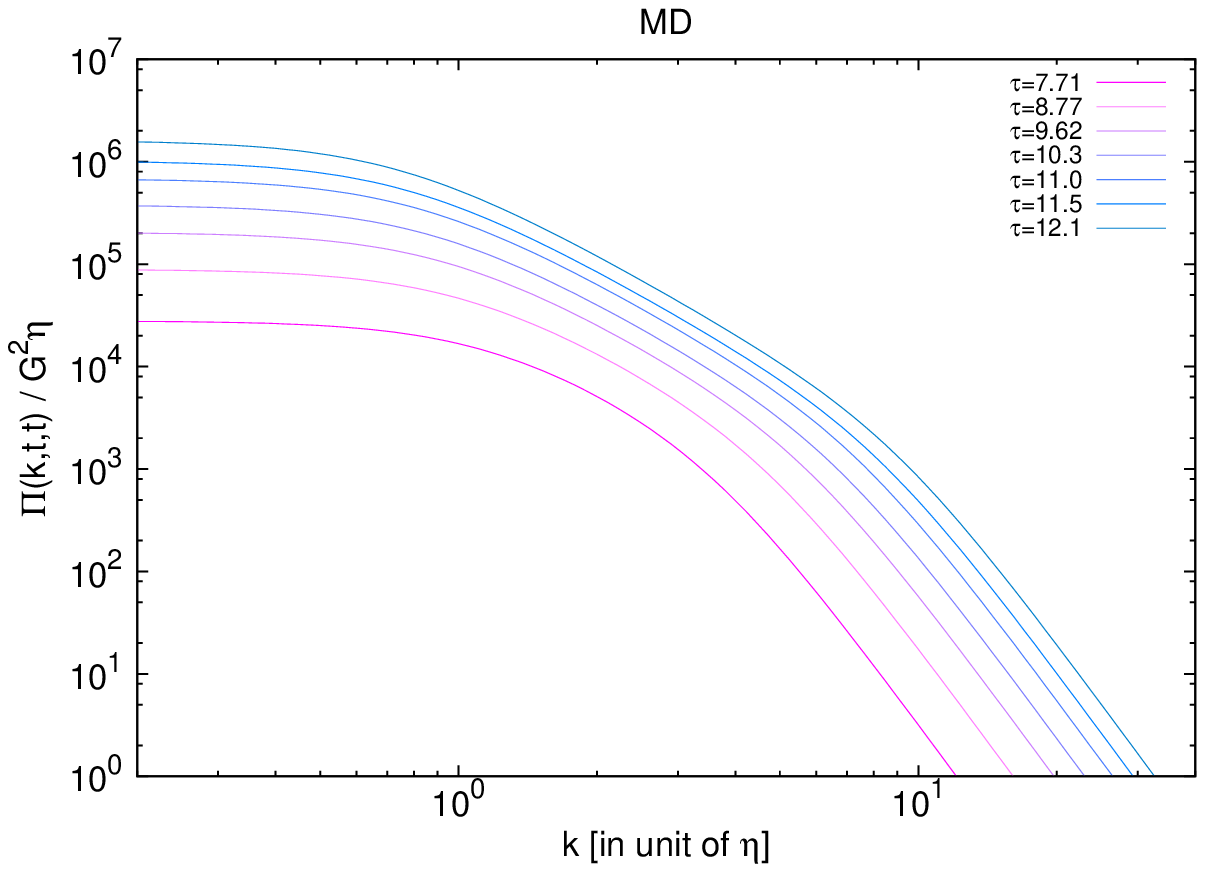}}
\end{array}$
\caption{The expression for $\Pi(k,\tau,\tau)$ given by eqs.~(\ref{eq4-9})-(\ref{eq4-12}) in
(a) radiation dominated background and (b) matter dominated background.}
\label{fig5}
\end{figure*}

We plot the function~(\ref{eq4-9}) in figure~\ref{fig5}. 
Comparing figures~\ref{fig4} and \ref{fig5}, we see that the function given by eqs.~(\ref{eq4-9})-(\ref{eq4-12})
indeed reproduces the form of $\Pi(k,\tau,\tau)$ obtained from numerical simulations
except that the spectrum in figure~\ref{fig5} does not agree with that in figure~\ref{fig4} at early times when domain walls
do not enter the scaling regime in the numerical simulation.
Fortunately, this disagreement does not significantly affect the final form of the GW spectrum because the value of $\Pi(k,\tau,\tau)$ at early times
is much less than that at the final time by a factor of ${\cal O}(10^{-4})$.

Now we have all the ingredients to evaluate the spectrum of GWs. The amplitude of GWs is given by eq.~(\ref{eq2-20}).
Setting $\beta=1/2$ and $\nu=1/2$ (in radiation dominated background), we obtain
\begin{align}
&(\Omega_{\mathrm{gw}})_* \notag\\
&= \frac{4}{3\pi^2}k^3\int^{x_f}_{x_i}\frac{dx_1}{x_1}\int^{x_f}_{x_i}\frac{dx_2}{x_2}\cos(x_1-x_2)\Pi(k,\tau_1,\tau_2), \label{eq4-13}
\end{align}
where the subscript $*$ represents the fact that it is not the spectrum of GWs at the present time but that in the radiation dominated era
(since the energy density of GWs is diluted as $\rho_{\mathrm{gw}}\propto a^{-4}$,
the ratio $(\Omega_{\mathrm{gw}})_*=\rho_{\mathrm{gw}}/\rho$ is independent of $t$ in the radiation dominated era).
Combining it with approximations given by eqs.~(\ref{eq4-3})-(\ref{eq4-5}), we can express the spectrum of GWs as a time integration of
the function $S(k,x)$ given by eq.~(\ref{eq4-10})
\begin{widetext}
\begin{equation}
(\Omega_{\mathrm{gw}})_*/\Omega_{\eta} = 
\left\{
\begin{array}{l l}
{\displaystyle \frac{1}{8\pi^3}\left(\frac{k}{\eta}\right)^3\frac{1}{x_{\eta}^7}\left\{
\left(\int^{x_f}_{x_i}dxx^{5/2}\cos x\sqrt{S(k,x)}\right)^2 + \left(\int^{x_f}_{x_i}dxx^{5/2}\sin x\sqrt{S(k,x)}\right)^2\right\}} & \mathrm{totally\ coherent}\\
{\displaystyle \frac{1}{8\pi^3}\left(\frac{k}{\eta}\right)^3\frac{\Delta x}{x_{\eta}^7}\int^{x_f}_{x_i}dx x^5 S(k,x)} & \mathrm{incoherent}\\
{\displaystyle \frac{1}{8\pi^3}\left(\frac{k}{\eta}\right)^3\frac{2}{x_{\eta}^7}\int^{x_f}_{x_i}dx x^6\int^{\tilde{x}}_x\frac{dy}{y}\cos(x-y)S(k,x)} & \mathrm{top\ hat},\\
\end{array}
\right. \label{eq4-14}
\end{equation}
\end{widetext}
where $x_{\eta}\equiv k\eta^{-1}$, $\Delta x\equiv k\Delta\tau$, $\tilde{x} \equiv \mathrm{min}\{x_f,x_c+x\}$, and $\Omega_{\eta}$ is given by eq.~(\ref{eq3-4}).

Similarly, setting $\beta=2/3$ and $\nu=3/2$ in matter dominated background, we obtain
\begin{widetext}
\begin{equation}
\Omega_{\mathrm{gw}}(\tau_f)/\Omega_{\eta} =
\left\{
\begin{array}{l l}
 \frac{1}{4\pi^2}\left(\frac{k}{\eta}\right)^3\frac{1}{x_f^2x_{\eta}^9}\left\{
\left(\int^{x_f}_{x_i}dxx^5N_{3/2}(x)\sqrt{S(k,x)}\right)^2 + \left(\int^{x_f}_{x_i}dxx^5J_{3/2}(x)\sqrt{S(k,x)}\right)^2\right\} & \mathrm{totally\ coherent}\\
 \frac{1}{4\pi^2}\left(\frac{k}{\eta}\right)^3\frac{\Delta x}{x_f^2x_{\eta}^9}\int^{x_f}_{x_i}dx x^{10}\left[(N_{3/2}(x))^2+(J_{3/2}(x))^2\right]S(k,x) & \mathrm{incoherent}\\
 \frac{1}{4\pi^2}\left(\frac{k}{\eta}\right)^3\frac{2}{x_f^2x_{\eta}^9}\int^{x_f}_{x_i}dx x^{19/2}\int^{\tilde{x}}_xdyy^{1/2}[N_{3/2}(x)N_{3/2}(y)+J_{3/2}(x)J_{3/2}(y)]S(k,x) & \mathrm{top\ hat}.\\
\end{array}
\right. \label{eq4-15}
\end{equation}
\end{widetext}
Note that, in the matter dominated era, $\Omega_{\mathrm{gw}}$ in eq.~(\ref{eq2-20}) depends on time.
In eq.~(\ref{eq4-15}), we fixed $\tau=\tau_f$, which means that eq.~(\ref{eq4-15}) represents the GW spectrum at the conformal time $\tau_f$.
This is the same quantity which we computed directly from lattice simulations in section~\ref{sec3-3} (see appendix~\ref{secA-2} for details).

In figure~\ref{fig6}, we show the spectrum of GWs evaluated from eqs.~(\ref{eq4-14}) and (\ref{eq4-15}) for each approximation, and also the spectrum
obtained by direct numerical calculations which we described in section~\ref{sec3-3}.
In this figure we fixed the values $\Delta\tau=1$ and $x_c=1$.
We found that the spectrum with top hat approximation has an agreement with
the result directly obtained from numerical simulations within a factor of ${\cal O}$(1).

\begin{figure*}[htp]
\centering
$\begin{array}{cc}
\subfigure[]{
\includegraphics[width=0.45\textwidth]{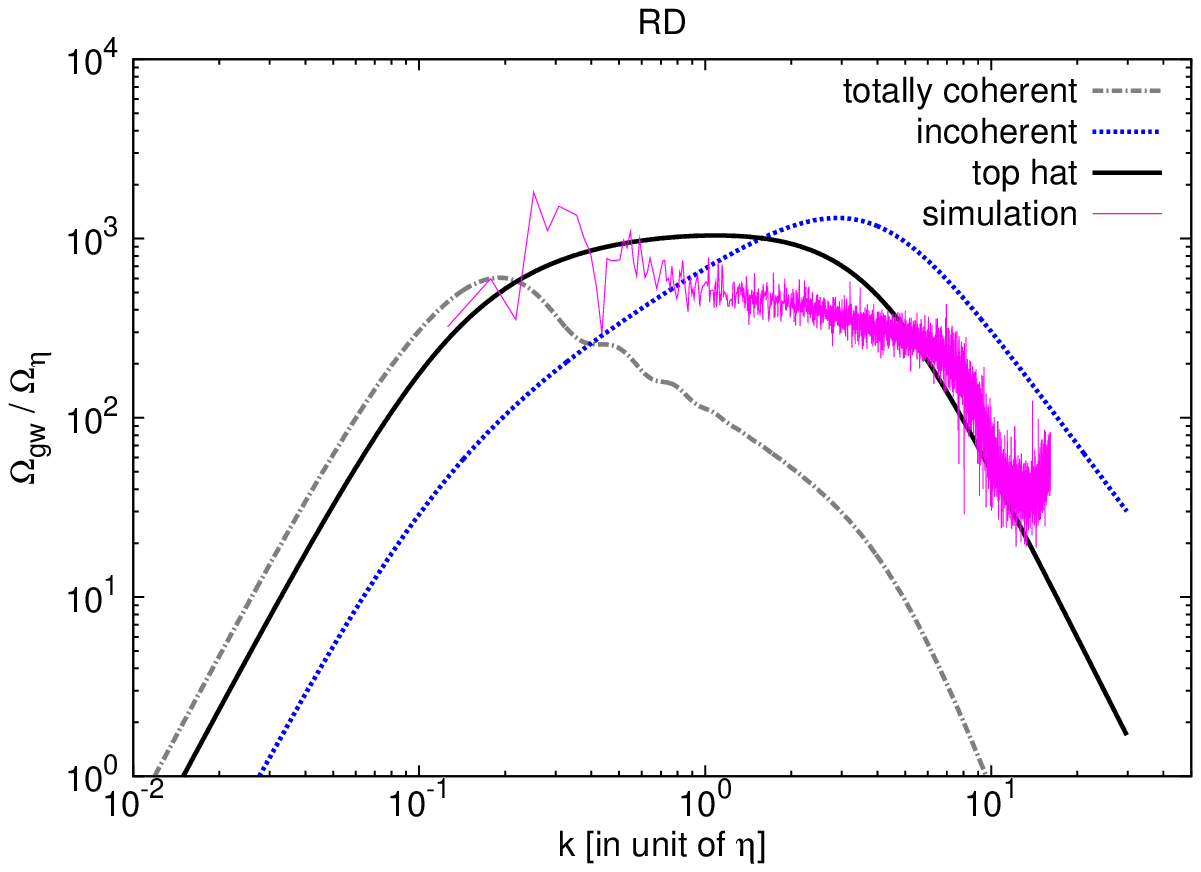}}
\hspace{20pt}
\subfigure[]{
\includegraphics[width=0.45\textwidth]{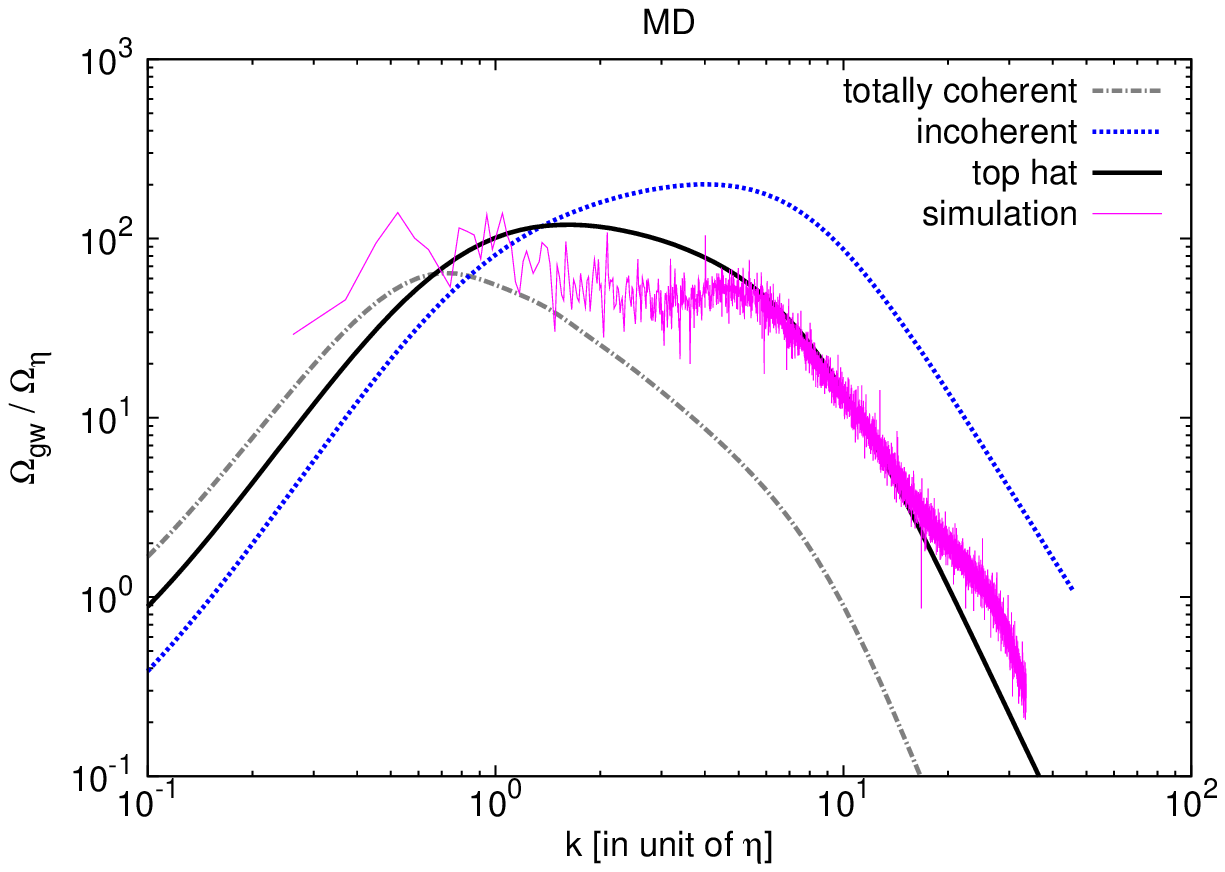}}
\end{array}$
\caption{The spectrum of GWs produced in (a) radiation dominated era and (b) matter dominated era, evaluated by using 
the totally coherent approximation (dot-dashed line), the incoherent approximation (dotted line), and the top hat approximation (thick solid line).
The thin solid line denoted by ``simulation" represents the spectrum directly obtained from numerical simulations.}
\label{fig6}
\end{figure*}

%%%%%%%%%%%%%%%%%%%%%%%%%%%%%%%%%%%%%%%%%%%%%%%%%%%%%%%%%%%%%%%%%%%%%%
\subsection{\label{sec4-3}Contribution from the time evolution of the source function}
%%%%%%%%%%%%%%%%%%%%%%%%%%%%%%%%%%%%%%%%%%%%%%%%%%%%%%%%%%%%%%%%%%%%%%
Now, let us examine the properties of approximations which we used in the previous subsections and consider their implications
for the shape of the GW spectrum.

Inspection of eqs.~(\ref{eq4-14}) and (\ref{eq4-15}) tells us that the $k$ dependence of $\Omega_{\mathrm{gw}}$
is determined by three factors: the phase space volume $k^3$, the time integral of the Green function and the function $S(k,\tau)$
in the equal time anisotropic stress power spectrum. In order to clarify the role of the individual terms, let us simply ignore the
time dependence of $S(k,\tau)$ and write the equal time anisotropic stress power spectrum as
\begin{equation}
\Pi(k,\tau,\tau) = G^2\eta f(\tau)S(k), \label{eq4-16}
\end{equation}
where $f(\tau)$ is a dimensionless function of $\tau$. In the previous subsection, we used $f(\tau)=(\tau\eta)^{\gamma}$
for domain walls, where $\gamma=7$ in radiation dominated background, and $\gamma=9$ in matter dominated background.
Substituting eq.~(\ref{eq4-16}) and approximations~(\ref{eq4-3})-(\ref{eq4-5}) into eq.~(\ref{eq2-20}), we obtain the following expression
\begin{equation}
\Omega_{\mathrm{gw}}(\tau_f)/\Omega_{\eta} = \frac{1}{8\pi^3}\left(\frac{k}{\eta}\right)^3F(k,\tau_i,\tau_f)S(k), \label{eq4-17}
\end{equation}
where $F(k,\tau_i,\tau_f)$ is given by
\begin{widetext}
\begin{align}
&F(k,\tau_i,\tau_f) = \frac{\pi\beta^2}{2(1-\beta)^2}x_{\eta}^{-\gamma}x_f^{\frac{2(1-2\beta)}{1-\beta}} \notag\\
&\qquad\qquad\qquad\times\left\{
\begin{array}{l l}
{\displaystyle \left[\left(\int^{x_f}_{x_i}dxx^{\frac{\beta}{1-\beta}-\frac{3}{2}+\frac{\gamma}{2}}N_{\nu}(x)\right)^2+\left(\int^{x_f}_{x_i}dxx^{\frac{\beta}{1-\beta}-\frac{3}{2}+\frac{\gamma}{2}}J_{\nu}(x)\right)^2\right]} & \mathrm{totally\ coherent} \\
{\displaystyle \Delta x\int^{x_f}_{x_i}dxx^{\frac{2\beta}{1-\beta}-3+\gamma}\left[\left(N_{\nu}(x)\right)^2+\left(J_{\nu}(x)\right)^2\right]} & \mathrm{incoherent}\\
{\displaystyle 2\int^{x_f}_{x_i}dxx^{\frac{\beta}{1-\beta}-\frac{3}{2}+\gamma}\int^{\tilde{x}}_{x}dyy^{\frac{\beta}{1-\beta}-\frac{3}{2}}\left[N_{\nu}(x)N_{\nu}(y)+J_{\nu}(x)J_{\nu}(y)\right]} & \mathrm{top\ hat.}
\end{array}
\right. \label{eq4-18}
\end{align}
\end{widetext}

\begin{figure*}[htp]
\centering
$\begin{array}{cc}
\subfigure[]{
\includegraphics[width=0.45\textwidth]{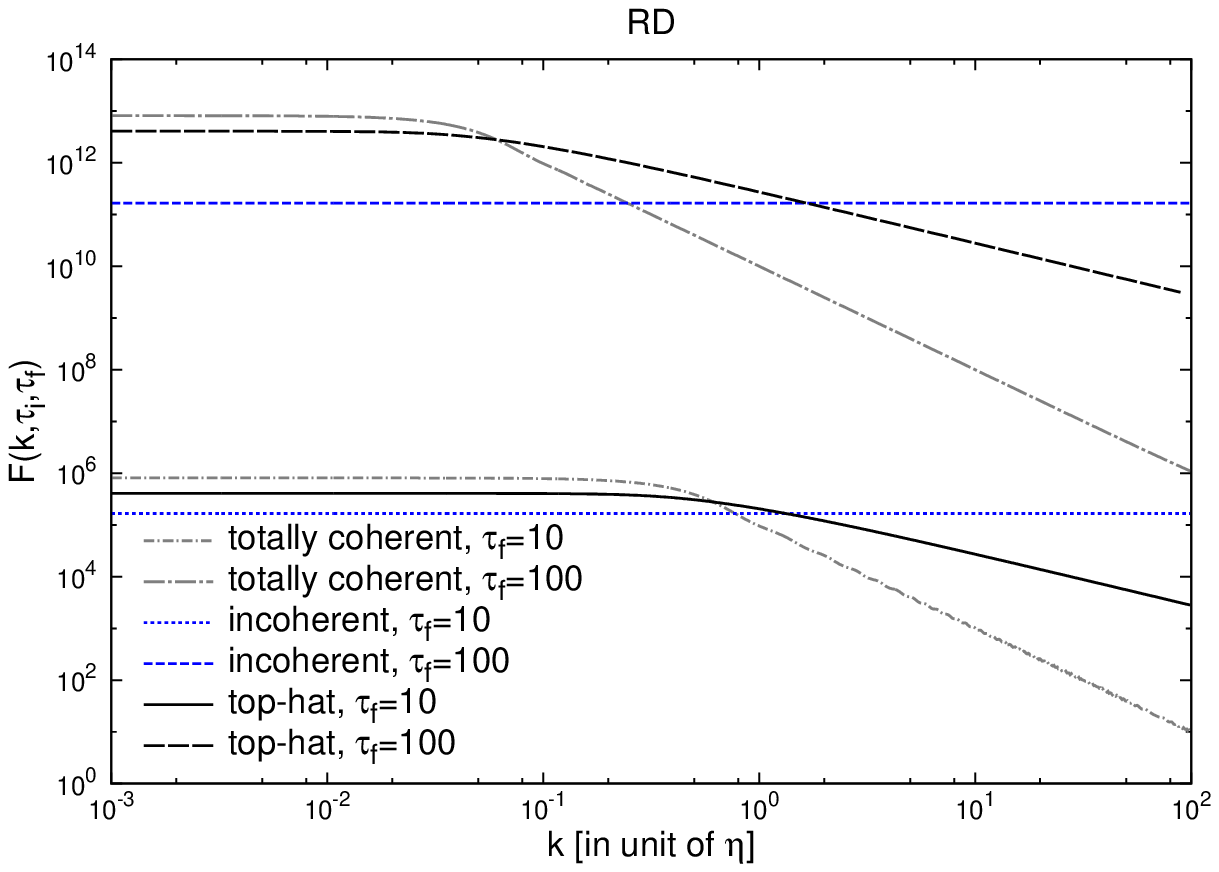}}
\hspace{20pt}
\subfigure[]{
\includegraphics[width=0.45\textwidth]{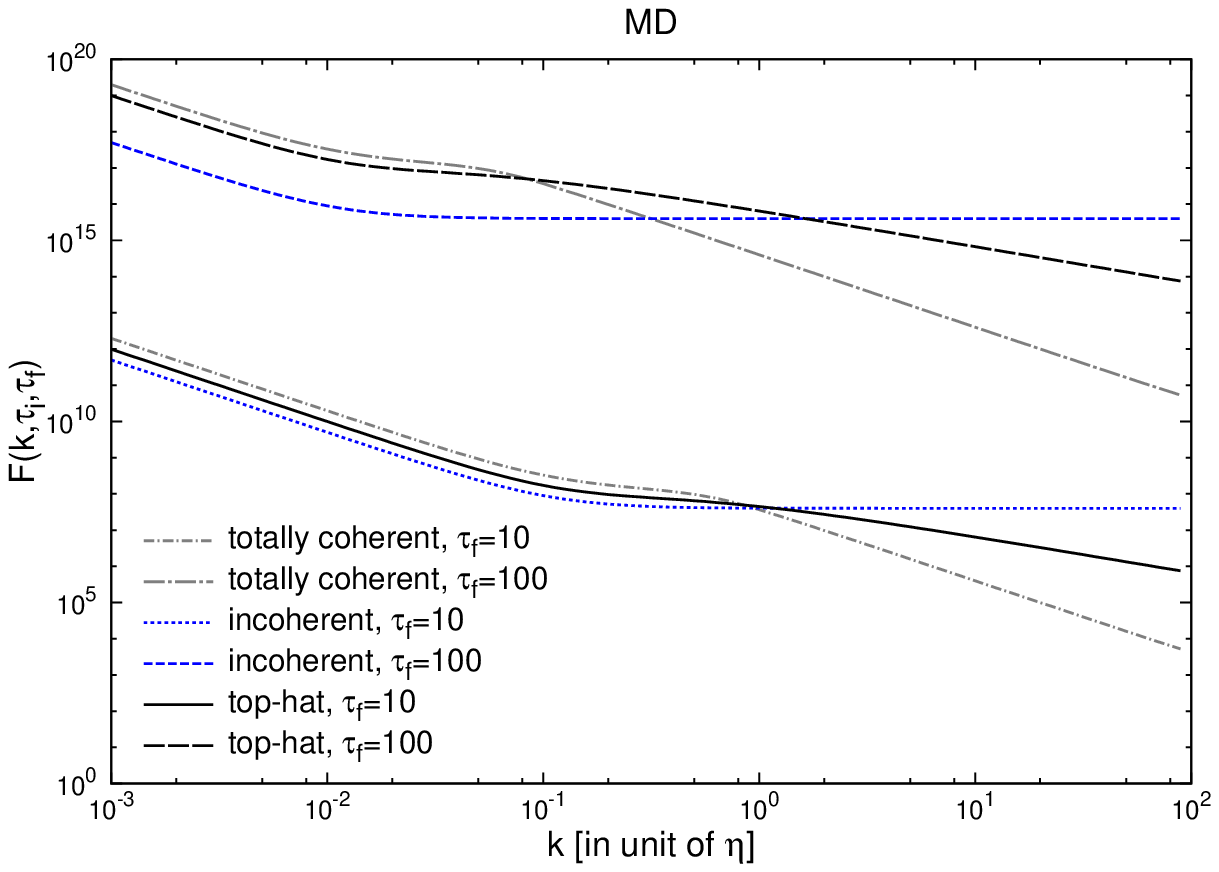}}
\end{array}$
\caption{The plot of $F(k,\tau_i,\tau_f)$ given by eq.~(\ref{eq4-18}) as a function of $k$.
We choose $\beta=\nu=1/2$ and $\gamma=7$ for radiation dominated background (a),
and $\beta=2/3$, $\nu=3/2$ and $\gamma=9$ for matter dominated background (b).
We computed for the case with $\tau_f=10\eta^{-1}$ and $\tau_f=100\eta^{-1}$ for each approximation.
Other parameters are fixed to be $\tau_i=2\eta^{-1}$, $\Delta\tau=1\eta^{-1}$ and $x_c=1$.}
\label{fig7}
\end{figure*}

In figure~\ref{fig7}, we plot $F(k,\tau_i,\tau_f)$ as a function of $k$ for each approximation.
This result shows that the factor $F(k,\tau_i,\tau_f)$
gives another contribution to the shape of the GW spectrum:
The factor $F(k,\tau_i,\tau_f)$ is suppressed for high frequencies $k\gtrsim \tau_f^{-1}$
in the totally coherent approximation and the top hat approximation,
while it is independent of $k$ in the incoherent approximation~\cite{2009JCAP...12..024C}.
This behavior is understood as a interference of the functions $N_{\nu}(x)$ and $J_{\nu}(x)$, which rapidly oscillate for the frequency
larger than $\tau_f^{-1}$. We find that $F(k,\tau_i,\tau_f)$ decays like $k^{-2}$ for the coherent case and $k^{-1}$ for the top hat case.
This means that in the totally coherent case the interference is stronger than that in the top hat case.
Therefore, there are much suppression at high frequencies for the totally coherent case, as we see in figure~\ref{fig6}.
On the other hand, the behavior of $F(k,\tau_i,\tau_f)$ at low frequencies $k<\tau_f^{-1}$ becomes different between the result with radiation dominated background
and that with matter dominated background. This is understood as a difference in the behavior of the Neumann function in the limit $x\to 0$.
Noting that $N_{\nu}(x)\propto x^{-\nu}$ for $x\ll 1$ and the integral in eq.~(\ref{eq4-18}) is dominated by the contribution around $x\approx x_i$,
one can easily show that $F(k,\tau_i,\tau_f) \propto k^{1-2\nu}$ for the modes with $k\ll \tau_f^{-1}$.
Therefore, in radiation dominated background ($\nu=1/2$), the amplitude of $F(k,\tau_i,\tau_f)$ does not depend on $k$ at lower frequencies.
This amplitude roughly scales as $\propto \tau_f^{\gamma}$ in the totally coherent approximation and the top hat approximation, 
and as $\propto \tau_f^{\gamma-1}\Delta\tau$ in the incoherent approximation.
This implies that the result with the incoherent approximation underestimates the GW amplitude than other two by a factor of $\Delta\tau/\tau_f$,
due to the presence of the additional factor $\Delta\tau$.
By contrast, in matter dominated background ($\nu=3/2$), $F(k,\tau_i,\tau_f)$ behaves as $k^{-2}$ at low frequencies $k<\tau_f^{-1}$.
This makes the tilt of the GW spectrum milder for the low frequencies than that in radiation dominated background, as we see in figure~\ref{fig6} (b).

\begin{figure}[htbp]
\includegraphics[width=0.45\textwidth]{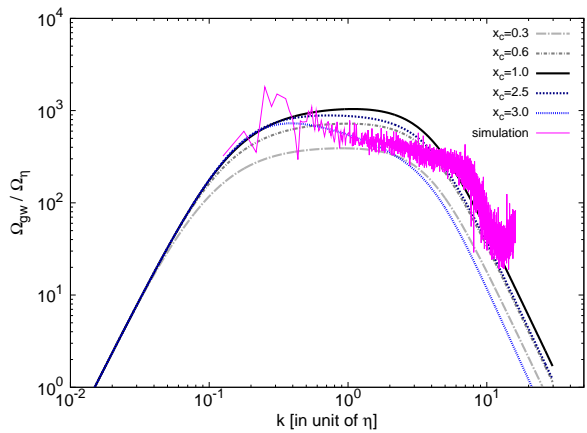}
\caption{The GW spectrum evaluated by using the top hat approximation with varying the value of $x_c$. The thin solid line
represents the spectrum directly obtained from numerical simulations.}
\label{fig8}
\end{figure}

In the top hat approximation, there is an additional parameter $x_c$, which can affect the shape of the GW spectrum.
In figure~\ref{fig8}, we show how the GW spectrum depends on the value of $x_c$.
Recall that the parameter $x_c$ determines the time interval beyond which the unequal time correlator vanishes.
If $x_c$ is large, the source is correlated for a long time and the amplitude of GWs is suppressed in the higher frequency modes
due to the presence of interferences.
This is caused by the integrand $\cos(x-y)$ in eq.~(\ref{eq4-14}), which has a dominant contribution for high frequencies
and gives a factor $\approx \sin(x_c)$. In particular, the higher frequency modes tend to vanish in the limit $x_c\to\pi$.
On the other hand, if $x_c$ is small, the whole amplitude of GWs is suppressed since the interval of integration in 
the third line of eq.~(\ref{eq4-14}) becomes short.
In this work we fixed $x_c=1$ as an intermediate value between two extreme cases described above.
We note that this choice may overestimate the amplitude of GWs compared with the result obtained directly from numerical simulations
in the intermediate scales between the Hubble radius and the width of walls.

The advantage of approximations which we have used is that we can predict the slope of the GW spectrum
for the frequencies which can not be resolved in the finite box lattice simulations.
Assuming the top hat case, the factor $F(k,\tau_i,\tau_f)$ in eq.~(\ref{eq4-17}) decays like $k^{-1}$ for the frequencies larger than $\sim\tau_f^{-1}$
in radiation dominated background. Note that the scale $\tau_f$ corresponds to the Hubble radius (in comoving coordinate) at the time $t_f$.
Therefore, the frequency at which the slope of $F(k,\tau_i,\tau_f)$ changes corresponds to $k_h(t_f)$ given by eq.~(\ref{eq3-5}).
Furthermore, from eq.~(\ref{eq4-10}) and table~\ref{tab1}, we expect that the factor $S(k)$ in eq.~(\ref{eq4-17}) is independent of $k$ for frequencies smaller than
$\sim k_h(t_f)$, while it decays like $k^{-2.02}$ for frequencies larger than $\sim k_h(t_f)$ and like $k^{-5.16}$ for frequencies larger than $\sim k_w(t_f)$.
Then, from eq.~(\ref{eq4-17}), we see that the tilt of the GW spectrum becomes $k^3$ for frequencies $k\lesssim k_h(t_f)$,
$k^3\cdot k^{-1}\cdot k^{-2.02}\propto k^{-0.02}$ for frequencies $k_h(t_f)\lesssim k\lesssim k_w(t_f)$ and
$k^3\cdot k^{-1}\cdot k^{-5.16}\propto k^{-3.16}$ for frequencies $k\gtrsim k_w(t_f)$.
The $k^{-0.02}$ behavior in the intermediate frequencies represents the slightly red tilted spectrum which we found from numerical simulations in section~\ref{sec3-3}.
However, it seems that the GW spectrum evaluated by using the top hat approximation in figure~\ref{fig6} does not decay like $k^{-0.02}$ for frequencies
$k_h(t_f)\lesssim k\lesssim k_w(t_f)$. This might be caused by the lack of the dynamical range.
As we see in next subsection, when these two scales $k_h(t_f)$ and $k_w(t_f)$ become separate by many order of magnitudes, the spectrum obtained by using this approximation
indeed gives the $k^{-0.02}$ behavior in the intermediate scales between the Hubble radius and the width of walls.

A similar reasoning can be applied to the case with matter dominated background.
In matter dominated background, assuming the top hat case,
$F(k,\tau_i,\tau_f)$ behaves like $k^{-2}$ for the frequencies smaller than $\sim \tau_f^{-1}$ and like $k^{-1}$ for
the frequencies larger than $\sim\tau_f^{-1}$. Combining this fact with the results shown in table I,
we expect that the tilt of the GW spectrum becomes $k^3\cdot k^{-2}\propto k$ for frequencies $k\lesssim k_h(t_f)$,
$k^3\cdot k^{-1}\cdot k^{-2.58}\propto k^{-0.58}$ for frequencies $k_h(t_f)\lesssim k\lesssim k_w(t_f)$ and
$k^3\cdot k^{-1}\cdot k^{-6.12}\propto k^{-4.12}$ for frequencies $k\gtrsim k_w(t_f)$.

%%%%%%%%%%%%%%%%%%%%%%%%%%%%%%%%%%%%%%%%%%%%%%%%%%%%%%%%%%%%%%%%%%%%%%
\subsection{\label{sec4-4}Spectrum today}
%%%%%%%%%%%%%%%%%%%%%%%%%%%%%%%%%%%%%%%%%%%%%%%%%%%%%%%%%%%%%%%%%%%%%%
The GW spectrum which we observe today is obtained by considering redshift due to the expansion of the universe.
Assuming that GWs are produced in the radiation dominated era, the amplitude of GWs observed today
is given by~\cite{1994PhRvD..49.2837K}
\begin{equation}
\Omega_{\mathrm{gw}}h^2(t_0) = 1.67\times 10^{-5}\left(\frac{100}{g_*}\right)^{1/3}(\Omega_{\mathrm{gw}})_*, \label{eq4-19}
\end{equation}
where $g_*$ is the number of relativistic degrees of freedom at the time when GWs are produced, $h$ is the renormalized Hubble
parameter ($H_0=100h$kmsec$^{-1}$Mpc$^{-1}$), and $(\Omega_{\mathrm{gw}})_*$ is given by eq.~(\ref{eq4-14})
(here we use the top hat approximation).  The frequency of GWs is given by
\begin{equation}
f(t_0) = \frac{k}{2\pi a_0} = \frac{k}{2\pi}\left(\frac{g_0}{g_*}\right)^{1/3}\frac{T_0}{T_i}, \label{eq4-20}
\end{equation}
where $a_0$ is the scale factor at the present time, $g_0=3.36$ is the number of relativistic degrees of freedom today, 
$T_0=2.725\mathrm{K}$ is the temperature of the universe observed today, and $T_i$ is the temperature at the time $t_i$.
In the second equality of eq.~(\ref{eq4-20}), we used $a(t_i)=1$, which is assumed in numerical simulations.
Recalling that $t_i$ is the time at which $t_i\eta=1$ is satisfied, we obtain
\begin{equation}
f(t_0) = 9.56\times10^8\times\left(\frac{100}{g_*}\right)^{1/12}\left(\frac{\eta}{10^{15}\mathrm{GeV}}\right)^{1/2}\left(\frac{k}{\eta}\right)\ \mathrm{Hz}. \label{eq4-21}
\end{equation}
The time when the production of GWs terminates is determined by the lifetime of the domain wall networks which decay due to the existence of
the bias term (\ref{eq3-3}). It is given by~\cite{1989PhRvD..39.1558G,2010JCAP...05..032H}
\begin{equation}
t_f = t_{\mathrm{dec}} = \frac{1}{2}\sqrt{\frac{\lambda}{2}}(\epsilon\eta)^{-1}. \label{eq4-22}
\end{equation}

Combining eqs.~(\ref{eq4-14}), (\ref{eq4-19}), (\ref{eq4-21}) and (\ref{eq4-22}), we can calculate the spectrum of GWs observed today.
Strictly speaking, it is inappropriate to put $t_i\eta=1$, because the occurrence of the domain wall networks would be much later if the phase transition
is driven by the finite temperature effect at $T\sim\eta$. However, it hardly affects the final shape of the GW spectrum if domain walls survived
for sufficiently long time, since the dominant contribution for the spectrum of GWs mainly come from GWs produced at the later time,
as we discussed in section~\ref{sec3-1}.

\begin{figure*}[htbp]
\centering
\includegraphics[scale=0.5]{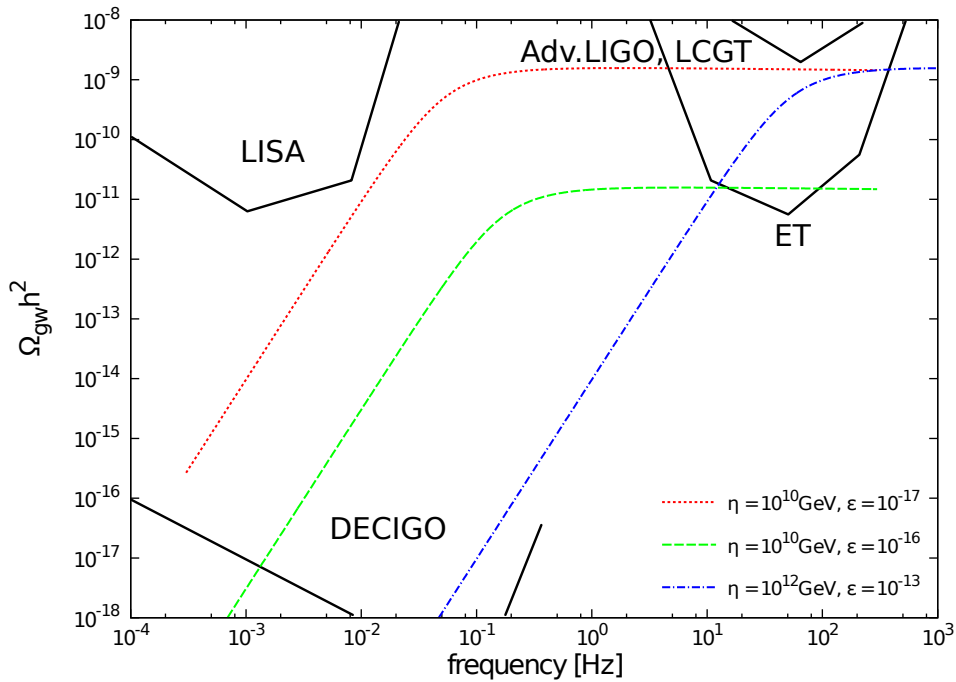}
\caption{The GW spectrum from domain walls for the case with ($\eta,\epsilon$) = ($10^{10}$GeV,$10^{-17}$) (dotted line),
($\eta,\epsilon$) = ($10^{10}$GeV,$10^{-16}$) (dashed line), and ($\eta,\epsilon$) = ($10^{12}$GeV,$10^{-13}$) (dash-dotted line).
Solid lines represent the rough sensitivity of planned detectors. Other parameters are chosen so that $\lambda=0.1$ and $g_*=100$.}
\label{fig9}
\end{figure*}

We show the result of the GW spectrum calculated by using the procedure described above in figure~\ref{fig9}.
We also show the expected sensitivity from future GW observations such as
Advanced LIGO~\cite{AdvLIGO}, LCGT~\cite{2002CQGra..19.1237K}, ET~\cite{ET}, LISA~\cite{LISA}, and DECIGO~\cite{2006CQGra..23S.125K}.
As we anticipated in~\cite{2010JCAP...05..032H}, GWs produced by domain walls with energy scale $\eta\approx10^{10}$GeV and sufficiently small $\epsilon$
are relevant to future GW direct detection experiments.
In particular, the frequency corresponding to the Hubble radius at the decay of domain walls, at which the tilt of the spectrum changes,
is located within the range of DECIGO and ground-based interferometers such as advanced LIGO, LCGT and ET.

We emphasize that there is a large change in dynamical range between the predicted spectra in figure~\ref{fig9} ($\epsilon\sim 10^{-13}$-$10^{-17}$)
and the spectra in figure~\ref{fig6} obtained from numerical simulations ($\epsilon\sim 10^{-3}$). We have to assume that the spectrum is flat enough
to extend over a large frequency range in order to give observable spectra which we plot in figure~\ref{fig9}.
Therefore, the results shown in figure~\ref{fig9} should be regarded as just an extrapolation of the flat spectrum obtained from numerical simulations.
This flatness property of the GW spectra must be confirmed by the future numerical studies with much larger dynamical range.

%%%%%%%%%%%%%%%%%%%%%%%%%%%%%%%%%%%%%%%%%%%%%%%%%%%%%%%%%%%%%%%%%%%%%%
\section{\label{sec5}Conclusions}
%%%%%%%%%%%%%%%%%%%%%%%%%%%%%%%%%%%%%%%%%%%%%%%%%%%%%%%%%%%%%%%%%%%%%%
In this work, we have computed the spectrum of GWs produced by domain walls based on the three dimensional lattice simulation of the scalar field.
For the evaluation of the GW spectrum, we apply two methods: One is to calculate the GW spectrum directly from numerical simulations
as in section~\ref{sec3-3}, and another is to calculate it indirectly by estimating the equal time anisotropic stress power spectrum and using an approximation for
the unequal time correlator as in section~\ref{sec4}. In the later case, we evaluate the GW spectrum according to the following procedure:
First, we compute the equal time anisotropic stress power spectrum as eq.~(\ref{eq4-2}). Then we assume the form of the power spectrum as
eq.~(\ref{eq4-9}), and determine the $k$-dependence of it by fitting with the result of the numerical computation as eq.~(\ref{eq4-10}).
For the unequal time correlator, we apply three approximations given by eqs.~(\ref{eq4-3})-(\ref{eq4-5}). Finally we obtain the GW spectrum by using the
formula~(\ref{eq2-20}). We find that the result with the top hat approximation given by eq.~(\ref{eq4-5})
agrees with the spectrum directly obtained from lattice simulations within a factor of ${\cal O}$(1).

From these analysises, we find the following features about the spectrum of GWs produced by domain walls:
\begin{enumerate}[(1)]
\item The slope of the spectrum changes at two characteristic frequencies corresponding to the Hubble radius at the decay of domain walls
and the width of domain walls. 
\item The spectrum between two characteristic frequencies described above becomes flat (or slightly red tilted) as $k^{-0.02}$ (in radiation dominated background).
\item Around the time when domain walls collapse, high frequency modes grow since the false vacuum regions fragment into small pieces (this effect is not included
in the analysis performed in section~\ref{sec4}).
\end{enumerate}

The indirect calculation performed in section~\ref{sec4} enables us to evaluate
the GW spectrum beyond the frequencies accessible in numerical simulations.
This method might be more effective than the direct numerical calculation for the problem which require
a long dynamical range, like domain walls.
However, we note that the results obtained in section~\ref{sec4} rely on various nontrivial assumptions:
First, the form of the function $S(k,\tau)$ given by eq.~(\ref{eq4-10}) should be regarded as a tentative
(i.e. other functions may also reproduce the form of
$\Pi(k,\tau,\tau)$ obtained from numerical simulations).
Second, although the top hat ansatz (\ref{eq4-5}) reproduces the spectrum obtained by numerical simulations,
there is no rigorous proof for the validity of this approximation.
Finally, in the calculation of the GW spectrum, we always assume that the source suddenly appears at the time $t_i$ and suddenly disappears at the time $t_f$.
In reality, the the source term $T_{ij}^{\mathrm{TT}}$ would gradually raise and continuously decay. This time dependence might affect the shape of the spectrum of GWs,
as discussed in~\cite{2009PhRvD..79h3519C}. 
Nevertheless, if the future numerical studies enable us to have an advanced understanding about the properties of the source,
this indirect method can be the alternative to numerical simulations for the evaluation of the GW spectrum.
Therefore, this work should be viewed as a first step toward the understanding of the spectrum of GWs produced by domain walls.
%%%%%%%%%%%%%%%%%%%%%%%%%%%%%%%%%%%%%%%%%%%%%%%%%%%%%%%%%%%%%%%%%%%%%%
%%%%%%%%%%%%%%%%%%%%%%%%%%%%%%%%%%%%%%%%%%%%%%%%%%%%%%%%%%%%%%%%%%%%%%

\begin{acknowledgments}
KS thank T. Hiramatsu for discussion on the numerical simulations.
This work is supported by Grant-in-Aid for
Scientific research from the Ministry of Education, Science, Sports, and
Culture (MEXT), Japan, No.14102004 and No.21111006 (M.K.)  and also by
World Premier International Research Center Initiative (WPI Initiative),
MEXT, Japan.
KS is supported by the Japan Society for the Promotion of Science (JSPS) through research fellowships.
\end{acknowledgments}

%%%%%%%%%%%%%%%%%%%%%%%%%%%%%%%%%%%%%%%%%%%%%%%%%%%%%%%%%%%%%%%%%%%%%%

\appendix

%%%%%%%%%%%%%%%%%%%%%%%%%%%%%%%%%%%%%%%%%%%%%%%%%%%%%%%%%%%%%%%%%%%%%%
\section{\label{secA-1}Formulation of the lattice simulations}
%%%%%%%%%%%%%%%%%%%%%%%%%%%%%%%%%%%%%%%%%%%%%%%%%%%%%%%%%%%%%%%%%%%%%%
The lattice formulation is similar to that used in our previous study~\cite{2010JCAP...05..032H} except the points which we described in section~\ref{sec3-1}.
In the numerical studies, we normalize the dimensionful quantities in the unit of $\eta$. For example, $\phi\to\phi/\eta$, $t\to t\eta$, etc.
With this normalization, we solve the equation of motion for the scalar field given by eq.~(\ref{eq3-1}) in the three dimensional lattice
by using the fourth order Runge-Kutta method. We put the periodic boundary condition in the configuration of the scalar fields.

The simulations are performed in the comoving box with size $b$ (in the unit of $\eta^{-1}$). The lattice spacing is $\delta x=b/N$,
where $N$ is the number of grid points (here we take $N=256$). We choose the initial time of the simulation so that $t_i=1$ and $a(t_i)=1$.
Then, the ratio of the Hubble radius to the physical lattice spacing $\delta x_{\mathrm{phys}}=a(t)\delta x$ is 
\begin{equation}
\frac{H^{-1}}{\delta x_{\mathrm{phys}}} = \frac{N}{b\beta}\left(\frac{t}{t_i}\right)^{1-\beta}, \label{eqA-1}
\end{equation}
and the ratio of the wall width $\delta_w=(\lambda^{1/2}\eta)^{-1}$ to the physical lattice spacing is
\begin{equation}
\frac{\delta_w}{\delta x_{\mathrm{phys}}} = \frac{N}{b\lambda^{1/2}}\left(\frac{t}{t_i}\right)^{-\beta}, \label{eqA-2}
\end{equation}
where $\beta$ is defined by eq.~(\ref{eq2-2}).
We take the final time and the size of the comoving box of the simulation as $t_f=151$, $b=50$ for the case with the radiation dominated background
and $t_f=65$, $b=24$ for the case with the matter dominated background. We take the value of coupling parameter as $\lambda=0.1$.
With these values of the parameters, at the end of the simulation, the ratios given by eqs.~(\ref{eqA-1}) and (\ref{eqA-2}) become
$H^{-1}/\delta x_{\mathrm{phys}}\simeq 125.8<N$, $\delta_w/\delta x_{\mathrm{phys}}\simeq 1.32$ for the case with radiation dominated background,
and $H^{-1}/\delta x_{\mathrm{phys}}\simeq 64.3<N$, $\delta_w/\delta x_{\mathrm{phys}}\simeq 2.09$ for the case with matter dominated background.
Therefore, these length scales are marginally resolvable even at the final time of the simulation.

We give the initial conditions so that the scalar field has quantum fluctuation at the initial time with correlation function in the momentum space given by
\begin{align}
\langle\phi({\bf k})\phi({\bf k'})\rangle = \frac{1}{2k}(2\pi)^3\delta^{(3)}({\bf k+k'}), \notag\\
\langle\dot{\phi}({\bf k})\dot{\phi}({\bf k'})\rangle = \frac{k}{2}(2\pi)^3\delta^{(3)}({\bf k+k'}). \label{eqA-3}
\end{align}
We used the massless fluctuations as the initial conditions, since the scalar field is near the top of potential barrier between two vacua at the initial time.
We also put the momentum cutoff $k_{\mathrm{cut}}$ above which all fluctuations are set to zero in order to eliminate the unphysical noise which comes from
high frequency modes in the field distributions. Here we set $k_{\mathrm{cut}}=1$.
We generate initial conditions in momentum space as Gaussian random amplitudes satisfying eq.~(\ref{eqA-3}), then Fourier transform them into
the configuration space to give the spatial distribution of the field.
When we adopt these initial conditions, the field distribution in the momentum space is dominated by the modes around $k \sim 1$
(or $k \sim \eta$, since we normalize all dimensionful quantities in the unit of $\eta$).
This means that, in the real space, the field value varies with a characteristic length scale $L \sim \eta^{-1}$.
This length scale is comparable to the Hubble radius at the initial time, since we take $t_i = 1$ (in the unit of $\eta^{-1}$).
Therefore, we expect that these initial conditions are likely to lead scaling domain wall configurations which satisfy the property $L \sim t$.
However, we emphasize that these initial conditions are chosen just for the convenience of the numerical study.
We use them to remove some difficulties that we find when we use the thermal initial conditions (see section~\ref{sec3-1}).
Since it seems that the scaling property is not so much affected by the initial field configuration and we are interested
in the evolution of the field after the formation of scaling domain wall networks, we expect that the results were qualitatively unchanged if we used the different initial conditions.

For the calculation of the area density of domain walls shown in figure~\ref{fig1}, we use the same algorithm as we used in~\cite{2010JCAP...05..032H}.

%%%%%%%%%%%%%%%%%%%%%%%%%%%%%%%%%%%%%%%%%%%%%%%%%%%%%%%%%%%%%%%%%%%%%%
\section{\label{secA-2}Direct calculation of gravitational waves} 
%%%%%%%%%%%%%%%%%%%%%%%%%%%%%%%%%%%%%%%%%%%%%%%%%%%%%%%%%%%%%%%%%%%%%%
In this appendix, we describe the method which is used in section~\ref{sec3-3} to calculate the spectrum of GWs from lattice simulations.

Instead of using the expression (\ref{eq2-17}), we replace the ensemble average in eq.~(\ref{eq2-16}) by an average over a volume $V$ of the comoving box,
\begin{equation}
\rho_{\mathrm{gw}} = \frac{1}{32\pi Ga^4}\frac{1}{V}\int\frac{d^3{\bf k}}{(2\pi)^3}\bar{h}'_{ij}(\tau,{\bf k})\bar{h}'^*_{ij}(\tau,{\bf k}). \label{eqB-1}
\end{equation}
Substituting the solution $\bar{h}_{ij}$ given by eqs.~(\ref{eq2-9}) and (\ref{eq2-10}), and ignoring the terms with higher order in $aH$, we obtain
\begin{align}
\rho_{\mathrm{gw}} =& \frac{2\pi^2G}{a^4V}\int\frac{d^3{\bf k}}{(2\pi)^3}\frac{1}{k^2} \notag \\
&\times \sum_{ij}\left\{ \left|\int^{x_f}_{x_i}dx'\sqrt{x'}a(x')N_{\nu}(x')T^{\mathrm{TT}}_{ij}(\tau',{\bf k})\right|^2 \right.\notag\\
&\qquad \left.+  \left|\int^{x_f}_{x_i}dx'\sqrt{x'}a(x')J_{\nu}(x')T^{\mathrm{TT}}_{ij}(\tau',{\bf k})\right|^2\right\}, \label{eqB-2}
\end{align}
where we used the approximation given by eq.~(\ref{eq2-18}), and averaged over a period of the oscillation of $\bar{h}_{ij}(\tau,{\bf k})$ with time.
The fraction of the energy density of GWs at the time $t$ given by eq.~(\ref{eq2-19}) becomes
\begin{align}
\Omega_{\mathrm{gw}}(t) =& \frac{2G^2k}{3Va(t)^4H(t)^2} \int d\Omega_k\notag \\
&\times \sum_{ij}\left\{ \left|\int^{x_f}_{x_i}dx'\sqrt{x'}a(x')N_{\nu}(x')T^{\mathrm{TT}}_{ij}(\tau',{\bf k})\right|^2 \right. \notag\\
&\qquad\left.+  \left|\int^{x_f}_{x_i}dx'\sqrt{x'}a(x')J_{\nu}(x')T^{\mathrm{TT}}_{ij}(\tau',{\bf k})\right|^2\right\}, \label{eqB-3}
\end{align}
where $\Omega_k$ is a unit vector representing the direction of ${\bf k}$ and $d\Omega_k=d\cos\theta d\phi$.
The TT part of the stress-energy tensor is computed by applying the projection operator in the momentum space
\begin{align}
T^{\mathrm{TT}}_{ij}(\tau,{\bf k}) &= \Lambda_{ij,kl}(\hat{k})T_{ij}(\tau,{\bf k}) \notag\\
&= \Lambda_{ij,kl}(\hat{k})\{\partial_k\phi\partial_l\phi\}(\tau,{\bf k}), \label{eqB-4}\\
\Lambda_{ij,kl}(\hat{k}) &= P_{ik}(\hat{k})P_{jl}(\hat{k}) - \frac{1}{2}P_{ij}(\hat{k})P_{kl}(\hat{k}), \label{eqB-5}\\
P_{ij}(\hat{k}) &= \delta_{ij}-\hat{k}_i\hat{k}_j, \label{eqB-6}
\end{align}
where $\hat{k}={\bf k}/|{\bf k}|$, and $\{\partial_k\phi\partial_l\phi\}(\tau,{\bf k})$ is the Fourier transform of $\partial_k\phi(\tau,{\bf x})\partial_l\phi(\tau,{\bf x})$.

Using the formulae described above, we can compute the GW spectrum as follows:
First, we obtain the time evolution of the scalar field $\phi(t,{\bf x})$ from the lattice simulation, then we compute the TT projected stress-energy tensor
as eqs.~(\ref{eqB-4})-(\ref{eqB-6}). Finally we perform the time integration in eq.~(\ref{eqB-3}) to obtain the spectrum.
Note that, as we mentioned in section~\ref{sec2}, $\Omega_{\mathrm{gw}}(t)$ given by eq.~(\ref{eqB-3}) does not
represent the value which would be observed today. In the numerical study, we compute the quantity $\Omega_{\mathrm{gw}}(t_f)$,
which represents the spectrum of GWs just after the production of them. In order to evaluate the spectrum observed today, one has to multiply
the dilution factor which caused by the expansion of the universe (see section~\ref{sec4-4}).
%%%%%%%%%%%%%%%%%%%%%%%%%%%%%%%%%%%%%%%%%%%%%%%%%%%%%%%%%%%%%%%%%%%%%%
%%%%%%%%%%%%%%%%%%%%%%%%%%%%%%%%%%%%%%%%%%%%%%%%%%%%%%%%%%%%%%%%%%%%%%

%%%%%%%%%%%%%%%%%%%%%%%%%%%%%%%%%%%%%%%%%%%%%%%%%%%%%%%%%%%%%%%%%%%%%%
\section{\label{secA-3}Curve fitting} 
%%%%%%%%%%%%%%%%%%%%%%%%%%%%%%%%%%%%%%%%%%%%%%%%%%%%%%%%%%%%%%%%%%%%%%
In section~\ref{sec4-2}, we fit the equal time anisotropic stress power spectrum $\Pi(k,\tau_f,\tau_f)$ obtained from numerical simulations
to the expression
\begin{equation}
S(k) = A\left[\left(1+\left(\frac{K_h(\tau_f)}{B}\right)^{C}\right)\left(1+\left(\frac{K_w(\tau_f)}{D}\right)^{E}\right)\right]^{-1}, \label{eqC-1}
\end{equation}
where $K_h(\tau_f)$ and $K_w(\tau_f)$ are given by eq. (\ref{eq4-7}).
The parameters ($A$, $B$, $C$, $D$, $E$) are chosen so that the quantity
\begin{equation}
\chi^2 = \sum_k\frac{1}{\sigma^2_{\ln\Pi(k)}}\{\ln\Pi(k)-\ln S(k)\}^2 \label{eqC-2}
\end{equation}
is minimized (hereafter we omit the time argument of $\Pi(k,\tau,\tau)$, assuming that the fitting is applied for the result with $\tau=\tau_f$).
$\sum_k$ in eq.~(\ref{eqC-2}) means to sum over all discrete values of $k$.
$\sigma_{\ln\Pi(k)}$ is the standard deviation of $\ln\Pi(k)$, which is given by
\begin{equation}
\sigma_{\ln\Pi(k)} = \frac{\sigma_{\Pi(k)}}{\Pi(k)}.  \label{eqC-3}
\end{equation}
Noting that $\Pi(k)$ is computed as an average over the direction of ${\bf k}$ [see eq.~(\ref{eq4-2})],
we estimate $\sigma_{\Pi(k)}$ as a standard deviation of $\Pi({\bf k})$ which is computed at the lattice point
on the shell with $|{\bf k}|=k$, 
\begin{equation}
\sigma^2_{\Pi(k)} = \frac{1}{N_k}\sum_{|{\bf k}|=k}\{\Pi({\bf k})-\Pi(k)\}^2, \label{eqC-4}
\end{equation}
where $N_k$ is the number of lattice point (in momentum space) at which $|{\bf k}|=k$ is satisfied.

We calculate $\chi^2$ given by eq.~(\ref{eqC-2}) with varying the parameters ($A$, $B$, $C$, $D$, $E$)
and find the optimal values which minimize the value of $\chi^2$. We also determine the errors of ($A$, $B$, $C$, $D$, $E$)
from a set of values at which $\chi^2$ deviates by 1 from its minimum.

\end{document}